\documentclass{aa}  

\usepackage{graphicx}
\usepackage{txfonts}
\usepackage[version=4]{mhchem}
\usepackage{hyperref}
\usepackage{lscape}             % to rotate a single page table, example in appendix.

\begin{document}

\title{Effective supernova dust yields from rotating and non-rotating stellar progenitors}

  \author{Koki Otaki
          \inst{1,2,3}
          \and
          Raffaella Schneider
          \inst{1,2,3,4}
          \and
          Luca Graziani
          \inst{1,2,3}
          \and
          Alessandro Bonella
          \inst{1}
          \and
          Stefania Marassi
          \inst{1}
          \and
          \\Marco Limongi
          \inst{2,5,6}
          \and
          Simone Bianchi
          \inst{7}
          }

   \institute{Dipartimento di Fisica, ``Sapienza'' Universit$\grave{a}$ di Roma, Piazzale Aldo Moro 5, I-00185 Roma, Italy\\
              \email{koki.otaki@uniroma1.it}
         \and
             INAF/Osservatorio Astronomico di Roma, Via Frascati 33, I-00040 Monte Porzio Catone, Italy
         \and
             INFN, Sezione Roma I, Dipartimento di Fisica, ``Sapienza'' Universit$\grave{a}$ di Roma, Piazzale Aldo Moro 2, I-00185, Roma, Italy
         \and
             Sapienza School for Advanced Studies, Viale Regina Elena 291, I-00161 Roma, Italy
         \and
             Kavli Institute for the Physics and Mathematics of the Universe (WPI), The University of Tokyo Institutes for Advanced Study, The University of Tokyo, Kashiwa, Chiba 277-8583, Japan
         \and
             INFN. Sezione di Perugia, via A. Pascoli s/n, I-06125 Perugia, Italy
        \and
             INAF/Osservatorio Astronomico di Arcetri, Largo E. Fermi 5, I-50125 Firenze, Italy        
             }

% \author[0000-0002-3406-3099]{Koki Otaki}

% \author[0000-0001-9317-2888]{Raffaella Schneider}

% \author[0000-0002-9231-1505]{Luca Graziani}

% \author[0000-0001-9018-4867]{Stefania Marassi}

% \author[0000-0002-3164-9131]{Marco Limongi}

% \author[0000-0002-9384-846X]{Simone Bianchi}

   \date{}

% \abstract{}{}{}{}{} 
% 5 {} token are mandatory
 
  \abstract
  {
  Supernovae (SNe) are believed to be the dominant sources of dust production at high redshift. However, the reverse shock generated by the interaction of the SN forward shock and the interstellar medium (ISM) can significantly reduce the mass of newly formed dust in SN ejecta.
  This study quantifies the mass, composition, and grain size distribution of surviving dust after the passage of the reverse shock using the \texttt{GRASHrev} model.
  Our analysis covers a grid of SN models with progenitor masses $13\,M_\odot \leq m_\star \leq 120\,M_\odot$, metallicity $-3\leq\text{[Fe/H]}\leq 0$, and rotation velocities $v=0$ and $300\,\mathrm{km\,s^{-1}}$. The SN explosions are assumed to occur in a uniform ISM with densities $n_\mathrm{ISM}=0.05,\,0.5,$ and $5\,\mathrm{cm^{-3}}$.
  We find that the larger grains $(\gtrsim 10 \,\mathrm{nm})$ are more resistant to destruction by the reverse shock, with amorphous carbon dominating the surviving dust mass in most models. 
  The surviving dust mass decreases with increasing ISM density. For non-rotating progenitors, the maximum mass of dust surviving the passage of the reverse shock is $\simeq 0.02 \,M_\odot$ released by SN explosions of a $120 \, M_\odot$ progenitor with $\text{[Fe/H]}=0$ in the ISM density $0.5\,\mathrm{cm^{-3}}$, corresponding to $\simeq 4\%$ of the initial dust mass before the passage of the reverse shock. Similarly, among the rotating progenitors, a maximum surviving mass fraction is $\simeq 5\%$ with a final dust mass $\simeq 0.03 \, M_\odot$ in $\text{[Fe/H]}=-1$ models.
  Although the reverse shock has a strong destructive impact, our results indicate that, on very short timescales, SNe can enrich the ISM with carbonaceous grains ranging in size from approximately 1 nm to 100 nm (up to $\simeq 1 \,\mathrm{\mu m}$ in non-rotating models). This is especially notable given the recent detection of the 2175 \AA \, UV extinction bump in galaxies at $z > 6$, suggesting the early presence of such dust.
  }


   \maketitle
%
%-------------------------------------------------------------------

\section{Introduction}

Understanding the origin of interstellar dust has a number of important astrophysical implications. The composition and size distribution of the grains affect their emission and extinction properties \citep{draine2003}. 
The dust content in the interstellar medium (ISM) controls the efficiency of H$_2$ formation and hence the rate of star formation in galaxies \citep{stahler2005}.
The presence of dust grains in low-metallicity star forming regions affects the gas fragmentation scales and hence the nature of the originating stellar populations \citep{schneider2003, schneider2006, schneider2012b, schneider2012a, omukai2005, 
marassi2014, chiaki2014, chiaki2015}. Finally, photo-electric heating due to dust grains is
critical to maintain the multiphase structure of the ISM \citep{wolfire2003}.

In recent years, deep millimeter observations of high-$z$ galaxies have
been able to detect their ISM dust continuum emission, indicating that rapid dust enrichment can occur already at $z > 6$, not only in luminous quasars \citep{valiante2014,calura2014,venemans2018} or Ultra Luminous Infrared Galaxies \citep{marrone2018, hashimoto2018}, but also in normal star forming systems, that are more representative of the bulk of the galaxy population at those redshifts \citep{watson2015, laporte2017, knudsen2017, hashimoto2018b, bowler2018, tamura2019}. A systematic analysis of the dust content in star-forming galaxies at $z > 4$ has recently come from two ALMA Large Programs, the ALPINE survey targeting galaxies at $4 \leq z \leq 6$ \citep{lefevre2020, pozzi2021}, and the REBELS survey \citep{bouwens2022, sommovigo2022, inami2022} targeting galaxies at $6 \leq z \leq 8$. Despite the uncertainties affecting the dust mass determination, when only one single far-infrared (FIR) continuum data point is available (see however \citealt{witstok2023, algera2024}), these observations provide evidence in support of the existence of a fast and efficient dust formation channel, in agreement with model expectations \citep{mancini2015, mancini2016, popping2017, vijayan2019, graziani2020, dayal2022, ferrara2022, dicesare2023, palla2024}.

On timescales of $\lesssim 30 - 40\,\mathrm{Myr}$,  core-collapse supernovae provide the dominant source of dust enrichment (see \citealt{schneider2024} for a recent review). On longer timescales, the most massive among stars in the Asymptotic Giant Branch start to contribute, but their importance depends on the star formation history, on the stellar initial mass function \citep{valiante2009, schneider2014} and on their mass- and metallicity-dependent dust yields \citep{zhukovska2008, ferrarotti2006, ventura2012a, ventura2012b, ventura2014, ventura2018, dellagli2014, dellagli2015, dellagli2016, dellagli2017, nanni2013, nanni2014, nanni2015, nanni2016}.

Core collapse Supernova (SN) dust yields have been computed by a number of studies, within the framework of standard nucleation theory \citep{kozasa1991, todini2001, nozawa2003, nozawa2010, schneider2004, bianchi2007, marassi2014, marassi2015, marassi2019}, or following a kinetic approach \citep{cherchneff2009, cherchneff2010, sarangi2013, sarangi2015, sluder2016, lazzati2016, brooker2022}. Despite the high efficiencies of grain condensation in SN ejecta, it has become clear that not all the newly formed dust makes its way to the ISM, due to the processing of newly formed grains in SN remnants \citep{dwek2005, bianchi2007, nozawa2007, nath2008, silvia2010, silvia2012, marassi2015, bocchio2016, biscaro2016, micelotta2016, martinez2019, kirchschlager2019, kirchschlager2020, slavin2020, kirchschlager2023}.
The destructive effect of the SN reverse shock generated by the interaction of the SN forward shock and the ISM is expected to reduce the mass of newly formed dust, to modify its size distribution, and the relative abundance of different dust species. It is therefore very important to quantify these effects for at least two reasons:\\
({\it i}) the ISM dust enrichment from SN is expected to be the dominant source of dust production at very high redshift and in chemically less evolved galaxies, with $M_{\star} \lesssim 2.5 \times 10^8 \, M_\odot$ at $z \geq 4$ (see e.g. \citealt{mancini2015, mancini2016, graziani2020, dicesare2023}); \\
({\it ii}) the size distribution and composition of surviving dust grains have important implications for the subsequent dust evolution in the ISM (see e.g. \citealt{hirashita2013}). Indeed, small grains will be more vulnerable to destruction by interstellar shocks \citep{nozawa2006,mcKee1989,jones2011,bocchio2014}, but their larger surface to volume ratio will enhance the efficiency of grain growth in the cold neutral medium \citep{weingartner1999, asano2013, asano2014, zhukovska2016, zhukovska2018}. The latter effect has been shown to be required by chemical evolution models with dust in order to reproduce the observed large dust masses \citep{rowlands2014, valiante2014, michalowski2015, mancini2015, popping2017, graziani2020} and dust scaling relations \citep{schneider2016, ginolfi2018, dicesare2023}.
Additionally, the grain size distribution impacts grain charging and the optical properties of the grains, with important implications for the propagation of radiation in the dusty medium and the physical properties of the multiphase interstellar medium \citep{glatzle2019, glatzle2022}.

In this paper, we use the \texttt{GRASHrev} model described in \citet{bocchio2016} to quantify the impact of the reverse shock on the dust yields computed by \citet{marassi2019} for the grid of SN models with progenitor masses ranging between $13$ and $120 \, M_{\odot}$, initial metallicity varying from $\text{[Fe/H]} = -3$ to $\text{[Fe/H]} = 0$ and two different values of the rotation rate $v = 0$ and $300\,\mathrm{km\,s^{-1}}$ \citep{limongi2018}. We explored SN explosions with a fixed-energy of $1.2 \times 10^{51}\,\mathrm{erg}$ expanding in uniform ambient media with constant densities of $n_{\rm ISM} \simeq 0.05, \,0.5$ and $5\, \mathrm{cm^{-3}}$. The results of these calculations are meant to provide a quantitative estimate of the {\it effective} dust yields (the mass of dust that each SN is able to inject in the ISM) for a large grid of SN models and ambient medium densities, to be used in chemical evolution models with dust to assess the importance of SNe as dust producers. 

The paper is organized as follows. In Section \ref{sec:previous} we summarize the main results of previous works. In Section \ref{sec:SNdustyields} we briefly describe the grid of SN dust yields computed by \citet{marassi2019} that we use as an input to the \texttt{GRASHrev} calculations. In Section \ref{sec:reverseshock} we recall the main features of the \texttt{GRASHrev} model \citep{bocchio2016} and we present the results of the calculations in Section \ref{sec:results}. Our main findings and their implications for early cosmic dust enrichment are discussed in Section \ref{sec:discussion}. Finally, in Section \ref{sec:conclusions} we draw our main conclusions.

\section{Previous works} \label{sec:previous}
There are still large uncertainties on the amount of newly formed dust in SN ejecta that survives the passage of the reverse shock (see the recent reviews by \citealt{micelotta2018} and \citealt{schneider2024}). Some studies have used analytic approximations to describe the expansion of the SN blast wave and the dynamics of the reverse shock, and have followed dust destruction by thermal and non-thermal sputtering \citep{nozawa2007, bianchi2007, micelotta2013, micelotta2016, bocchio2016}. The general conclusions of all these studies are that the fraction of dust that survives ranges between $\sim 20\%$ to less than a few \% depending on the density of the medium into which the remnant is expanding, the supernova explosion energy, as well as the composition and size distribution of the grains that originally formed in the ejecta. Hence, the results differ for different types of supernovae and likely depend on the metallicity and mass of the progenitor star. 

\citet{bocchio2016} compared their predictions with four SNe in the Milky Way and Large Magellanic Cloud (SN 1987a, Cas A, the Crab Nebula, and N49). For each individual SN, the model predictions are in good agreement with the presently observed dust mass. However, the largest dust mass destruction is predicted to occur between $10^3$ and $10^5 \,\mathrm{yr}$ after the explosions, while the oldest SN in the sample has an estimated age of $4800\,\mathrm{yr}$. Hence, these observations can only provide an upper limit to the mass of dust that will be injected into the ISM. 

A different approach has been undertaken by \citet{silvia2010, silvia2012} and more recently by \citet{martinez2019, kirchschlager2019, kirchschlager2020, kirchschlager2023, slavin2020}. These authors have used hydrodynamic simulations to investigate the destruction of newly formed dust by the SN reverse shock. \citet{silvia2010,silvia2012} used an idealized set-up of a planar wave impacting on a dense spherical ejecta clump (cloud-crushing problem) in which they have implanted a distribution of Lagrangian particles to represent a population of grains whose sizes and composition were taken from the computation of \citet{nozawa2007} (for a SN progenitor of $20\,M_{\odot}$ with unmixed ejecta). They found that the dominant factor determining the degree of grain destruction is the relative velocity between the reverse shock and the ejecta cloud, with a threshold shock velocity between $1000\text{--}3000\,\mathrm{km\,s^{-1}}$ above which grain survival drops considerably. Their results indicate the grains with initial radii less than $0.1\,\mathrm{\mu m}$ are often completely destroyed, while larger grains survive their interaction with the reverse shock. \citet{martinez2019} used a 3D hydrodynamic simulation to explore the expansion of SN ejecta in a pre-existing wind-driven bubble. In this case, a higher survival fraction is found because the SN remnant evolves into a lower density medium excavated by the wind. They assumed a single SN with a progenitor star of $60 \,M_{\odot}$ that produces $0.5 \,M_{\odot}$ of ejecta dust\footnote{No information is given on the adopted ejecta grain population.}  and
found that if the expansion occurs in a uniform ambient medium with density $n = 1 \, {\rm cm}^{-3}$
($n = 1000 \, {\rm cm}^{-3}$) the dust survival fraction is $< 32\%$ (zero). However, with a pre-existing wind-driven bubble in the same ambient media the survival fractions increase to 
$\sim 95\%$. The crucial parameter here is the mass ratio between the mass of the wind-driven shell and the mass of the ejecta, which they have assumed to be $\chi = 400$ ($\chi = 2 \times 10^4$). In all these studies, however, perfect coupling between dust and gas is assumed and they do not consider the effects of non-thermal sputtering and shattering by grain-grain collisions. \citet{kirchschlager2019} used 2D hydrodynamics simulation to follow the interactions between grains and gas particles mediated by sputtering and grain-grain collisions in a cloud-crushing problem, showing that the probabilities of grain-grain collisions can be enhanced if the grains reside in over-dense clumps within the ejecta, although this effect is partly counteracted by grain growth \citep{kirchschlager2020}. In addition, a significant lower fraction of dust survives when including the effects of the magnetic field on the dynamics of charged grains, as the betatron acceleration can cause kinetic decoupling between gas and dust, enhancing grain sputtering \citep{slavin2015}. However, the effect of the magnetic fields depends on its morphology, as a uniform field has been shown to trap the grains, which are reflected back onto the remnant \citep{fry2020}, while a turbulent field allows the diffusion of the grains in the ISM \citep{slavin2020}.

Finally, we shall note that here we are focusing on the effective dust yields released by individual supernova explosions. Sequential supernova explosions originating from a coeval massive star cluster embedded in a clumpy molecular cloud may lead to different resulting dust masses, due to the partial destruction of SN-injected grains by subsequent explosions. On timescales of $\simeq 2.5\,\mathrm{Myr}$, this can be partly compensated by dust grain growth in the emerging super-bubble swept up super-shell, as shown by \citet{martinez2022}. While these effects are clearly important to consider when modeling the time evolution of the dust mass in galaxies, here we limit our study to estimating the mass, composition and size distribution of grains released by individual stellar explosions.

\section{The grid of SN dust yields}
\label{sec:SNdustyields}

The grid of SN dust yields that we use as an input to the reverse shock calculations has been computed and thoroughly described in \citet{marassi2019}. Here we briefly summarize the main features that are relevant to the reverse shock calculations, while we refer the reader to the original paper for more details.

The mass, composition, and size distributions of dust grains formed in SN ejecta have been computed using Classical Nucleation Theory. We adopted an improved version of the model of \cite{bianchi2007} that accounts for the non-steady-state formation of important molecular species \citep{marassi2014,marassi2015}. 
In particular, while the ejecta expands, we follow the formation and
destruction rates of CO, SiO, C$_2$, and O$_2$ molecules,
which play an important role in catalyzing or suppressing some
reaction rates that lead to dust formation.
We have computed the nucleation and accretion of seven different 
grain species: amorphous carbon (AC), iron (Fe), corundum (Al$_2$O$_3$),
magnetite (Fe$_3$O$_4$), enstatite (MgSiO$_3$), forsterite (Mg$_2$SiO$_4$)
and quartz (SiO$_2$).

Grain condensation depends on the ejecta temperature 
and density, whereas grain composition depends on the relative
abundances of refractory elements, which, in turn, 
depend on the nature of the SN progenitor 
(mass, metallicity, rotation) and on the explosion energy (see also \citealt{schneider2024}).

\subsection{SN models}
The pre-SN evolution is simulated with the \texttt{FRANEC} code
\citep{chieffi2013,limongi2018}, taking into account
the effects of stellar rotation and metallicity. 
The model grid spans a range of initial stellar masses
$[13\text{--}120]\, M_{\odot}$, that are assumed to explode 
with a fixed-energy of $1.2 \times 10^{51}\,\mathrm{ergs}$ (see for more
details the description of fixed-energy models in \citealt{marassi2019}). 
We assume two possible initial rotation rates of the stars, $v = 0$ (non-rotating models) 
and $v = 300\,\mathrm{km\,s^{-1}}$ (rotating models). 
Finally, we also consider 4 different initial metallicity values:
$\text{[Fe/H]} = 0$, $-1$, $-2$, and $-3$.
A thorough description of the properties of the SN models can be found in \citet{chieffi2013} and \citet{limongi2018}. 

Tables 1--4 in \cite{marassi2019} report the main properties of the SN models, including a classification of their type\footnote{We follow \citet{Hachinger2012} and classify the SN as IIP, IIb, or Ib depending on the H and He envelope masses in the pre-SN phase \citep{marassi2019}.}, and Tables 9 and 10 present their associated total dust yields. At the high progenitor mass-end, some of the entries in these Tables are missing either because the stars fail to explode (the ejecta do not contain enough material to trigger dust formation) or because their He cores are large enough to trigger pair production, the stars become pulsationally unstable and then their evolution can no longer be properly followed with \texttt{FRANEC} \citep{limongi2018}.

Here we report in the Appendix the initial dust mass in all grain species predicted for each of the FE SN models. In the next two subsections, we briefly summarize the main characteristics of this data. 

The mass of dust formed in the ejecta depends on the metal abundances and on their spatial distribution.
In the pre-SN phase, the models are characterized by a large variety of chemical species. However,
during the explosion, the mass of the ejecta can be significantly reduced due to the effect of fallback, which is particularly important for the most massive progenitors, that are characterized by larger binding energies. As a result, the mass of dust produced in the ejecta decreases for progenitor stellar masses above $\sim 25 \text{--} 30 \,M_\odot$ but in a non-monotonic way.

\subsection{Dust yields for non-rotating models}

For non-rotating stellar progenitors, the vast majority of the SN explosions are of type IIP and Ib.
The mass of dust produced depends on the progenitor mass and metallicity. The most efficient dust producers are SN models with progenitor masses $m_\star \sim 20 \text{--} 25 \,M_\odot$ that are able to synthesize up to $m_{\rm dust} \sim 1.2 \,M_\odot$ of dust. For $m_\star < 20 \text{--} 25 \,M_\odot$, the dust mass increases with progenitor mass, while for $m_\star > 20 \text{--} 25 \,M_\odot$ it drops due to the effect of fallback, reaching
values of $m_{\rm dust} \sim 0.2 \text{--} 0.4 \,M_\odot$ for the higher metallicity models ($\text{[Fe/H]}=0$ and $-1$) and dropping to $m_{\rm dust} \sim 0.01 \text{--} 0.02 \,M_\odot$ for models $\text{[Fe/H]}=-2$. At even lower metallicity, no dust is formed in SN ejecta with progenitors larger than $30 \,M_\odot$ (see Fig. 8 in \citealt{marassi2019} and the top solid lines in Fig. \ref{fig:totaldust_nonrot}).

The dust composition is very sensitive to the metallicity and mass of the stellar progenitors (see Fig. 10 in \citealt{marassi2019}). At higher metallicity ($\text{[Fe/H]}=0$ and $-1$), a variety of dust species form in the SN ejecta with progenitor masses $m_\star < 30\, M_\odot$, while SN with larger progenitors form mostly AC grains. At lower metallicity ($\text{[Fe/H]}=-2$ and $-3$), silicates can only form in a very limited number of SN models (those with progenitor masses in the range $20 \text{--} 25 \,M_\odot$), while in all the other models the dominant grain species are AC and Fe$_3$O$_4$.

The total dust mass and the mass of individual dust species for non-rotating SN models with different initial mass and metallicity are reported in Table \ref{tab:sndust_initial_non_rotating}.

\subsection{Dust yields for rotating models}

The effects of rotation-induced mixing are to increase the mass loss during the pre-SN evolution and to favour the formation of larger He and CO cores. Therefore, for the same initial stellar mass and metallicity, rotating models are characterized by a smaller mass at the onset of the SN explosion compared to non-rotating ones (see Tables 1-4 in \citealt{marassi2019}). As a result, except for the smallest initial stellar masses ($m_\star < 30 \, M_\odot$) with $\text{[Fe/H]}\leq -1$, all rotating models explode as type Ib SNe. However, rotation-induced mixing also leads to more metal-enriched ejecta and all rotating models with $\text{[Fe/H]} = 0, -1$ are able to form dust in their ejecta (there are no failed SN models). Within the grid of models used for the present analysis, the upper mass limit for a successful dust-producing SN is set by the onset of pair-production and of the associated pulsational instability that is experienced by stars with initial $\text{[Fe/H]}= -2$ and $-3$ and initial masses $m_\star \geq 80 \, M_\odot$ due to their larger He cores compared to non-rotating models. While we can not follow the evolution of pulsationally-unstable stars with FRANEC, we shall note that pair-instability supernovae are expected to produce large amount of dust, as shown by \citet{nozawa2003}, \citet{schneider2004}, and \citet{cherchneff2009, cherchneff2010}.

The total dust mass produced by rotating models ranges from a minimum value of 
$m_{\rm dust} \simeq 2.6 \times 10^{-3} \, M_\odot$ to a maximum value of $2.2 \, M_\odot$
with a strongly non monotonic behaviour with initial stellar mass and metallicity (see the top solid lines in all the panels of Fig. \ref{fig:totaldust_rot}). 

Similarly to non-rotating models, rotating stars with initial masses
$< 40 \, M_\odot$ (for $\text{[Fe/H]} = 0$) and $< 20 \, M_\odot$ (for $\text{[Fe/H]} = -3$) are able to synthesize a variety of grain species during their SN explosions. Conversely, SN with more massive progenitors mostly form AC grains. However, independent of metallicity, rotation leads to a more efficient formation of silicates and to a less efficient formation of AC grains in low-mass progenitors. Indeed, the physical conditions present in the ejecta of rotating models promote the formation of SiO and CO molecules that, in turn, affect the relative abundance of silicates and AC grains. 

The total dust mass and the mass of individual dust species for rotating SN models with different initial mass and metallicity are reported in Table \ref{tab:sndust_initial_rotating}.

\subsection{Grain size distributions}
It is important to stress that - for both rotating and non-rotating models - the grain size distribution is also very sensitive to the SN progenitor properties (as an example, see Fig. \ref{fig:distri_M20V0}). In general, the size distribution has a log-normal or bimodal shape, but the amplitude and peak size depend on the density at the onset of nucleation. Hence, SN models where the adiabatic expansion of the ejecta starts at smaller radii tend to form larger grains. Similarly, for a given SN model, the dust species whose nucleation starts earlier in the ejecta evolution (because of the larger dust nucleation temperature and/or higher concentration of key species in the gas phase), such as AC grains, are characterized by larger grain sizes. This has important consequences for grain survival during the passage of the reverse shock, as we will in the next sections.

\section{SN reverse shock modeling}
\label{sec:reverseshock}

In this section, we briefly summarize the main features of the \texttt{GRASHrev} model \citep{bocchio2016}, that we will use to quantify the impact of the reverse shock on the mass, composition and size distribution of dust grains produced by the SN models described in Section \ref{sec:SNdustyields}.

The evolution of the SN remnant into a uniform ambient medium is described by the unified, self-similar solution of \citet{truelove1999}, that fits a spherically symmetric, non-radiative hydrodynamical simulation from the Ejecta Dominated (ED, when the ejecta is in free-expansion) to the Sedov-Taylor (ST, when the mass of the interstellar material swept up by the shock wave becomes comparable to the mass of the ejecta) phases. The solution provides the position (with respect to the center of the explosion) and velocity of the forward and reverse shocks, but it breaks down when the velocity of the forward shock drops below $\sim 200\,\mathrm{km\,s^{-1}}$, and the remnant enters the Pressure-driven Snowplough (PDS) phase, where radiative losses can no longer be neglected. An analytic solution for this final phase is provided by \citet{truelove1999} and appears to be in good agreement with the results of the radiative hydrodynamic simulations of \citet{cioffi1988} and \citet{slavin2015}. 

The ejecta and the surrounding ISM (out to a distance of $200\,\mathrm{pc}$ from the center of the ejecta) are divided in $2 \, N_s=800$ spherical shells. Before
being crossed by a shock, the ejecta shells are assumed to be in homologous expansion and their temperature decreases adiabatically, while the ISM shells are assumed to be at rest. The timescales of the simulation are set by the time $\Delta t$ required by the reverse shock 
to cross one shell. When a shell is crossed by a shock, its temperature, density and velocity change according to the Rankine-Hugoniot jump conditions (see \citealt{bocchio2016} for additional details).

Initially, in the un-shocked ejecta, dust and gas are assumed to be uniformly mixed and perfectly coupled (they expand at the same velocity). However, when a given shell is hit by the reverse shock, the gas and dust respond differently, they decouple, and the grains time-dependent velocity is reduced according to the combined effects of collisional and plasma drag\footnote{We do not consider the effect of the betatron acceleration, that increases the velocity of charged grains in the presence of a magnetic field. This process is likely to be negligible inside spherically-symmetric ejecta, because both the magnetic field lines and the charged grains velocities are radially oriented. It is also negligible in the shocked ISM as during the PDS phase, when shock compression is important, the grains are almost at rest with the post-shock gas \citep{bocchio2016}. See also section \ref{sec:previous}.}. The model does not explicitly
compute the time-dependent ionization of the gas, but it assumes that
the shocked gas is fully ionized and composed by \ion{H}{2}, \ion{He}{3} and \ion{O}{2}, that
represent the most abundant ions.

Dust reprocessing is considered following thermal and non thermal sputtering, shattering and vapourisation in grain-grain collisions. 
The non-thermal sputtering term due to interactions with gas particles is described as in \citet{bocchio2014} using the sputtering yields adopted by \citet{nozawa2006} and \citet{bianchi2007}. Thermal sputtering that
causes sublimation due to collisional heating to high temperatures is described as in \citet{bianchi2007} and it is always found to be negligible. Shattering and vapourisation in grain-grain collisions are described as in \citet{jones1996}. For a detailed description we refer the reader to Appendix B and C of \citet{bocchio2016}.

Given the low-density of dust in the ejecta, grain-grain collisions are rare and shattering and vapourisation are always sub-dominant with respect to non-thermal sputtering, which represents the dominant dust destruction processes in SN ejecta.

\section{Results}
\label{sec:results}

\begin{figure*}
\includegraphics[width=.5\linewidth]{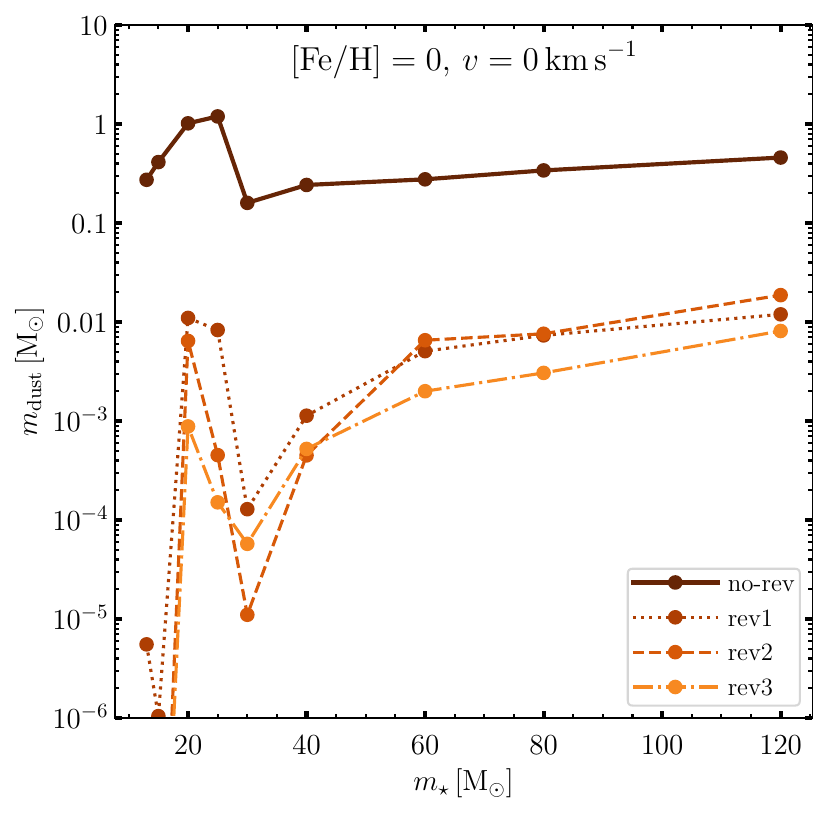}
\includegraphics[width=.5\linewidth]{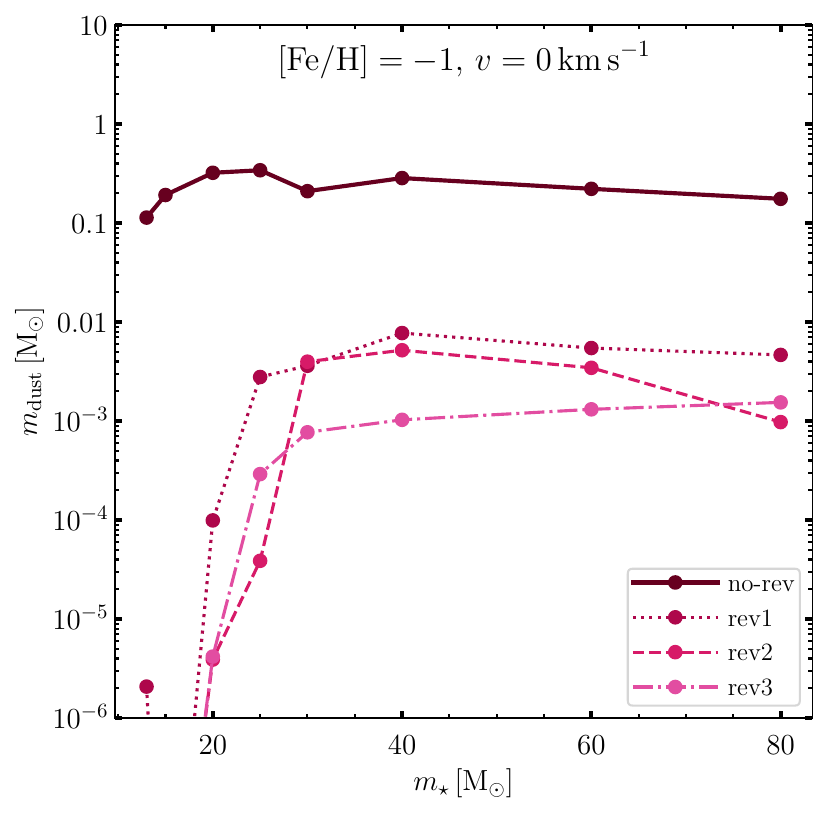}
\includegraphics[width=.5\linewidth]{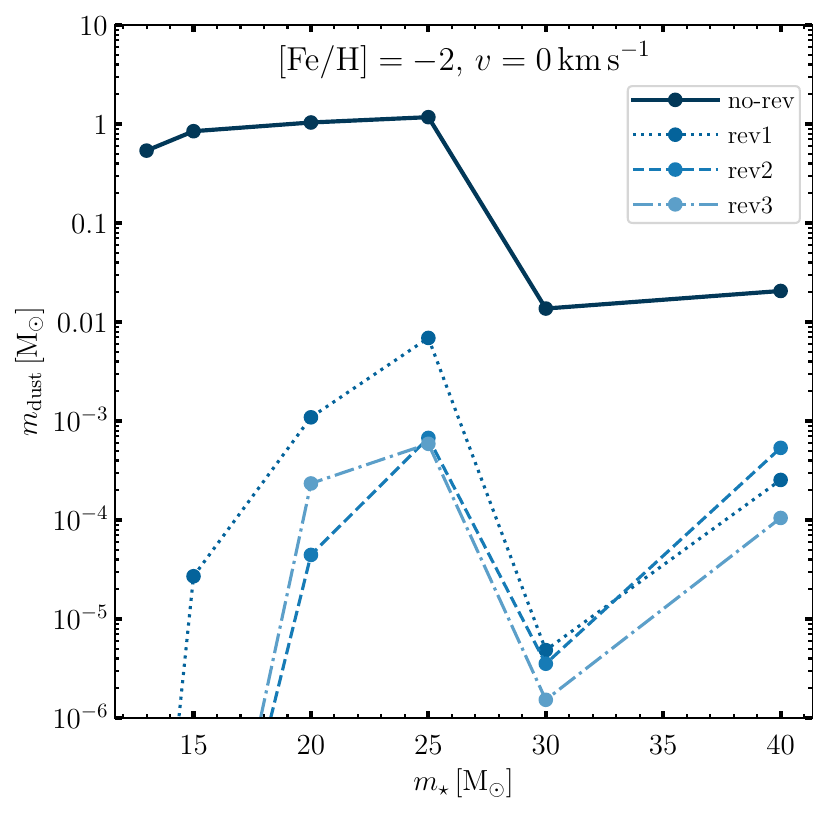}
\includegraphics[width=.5\linewidth]{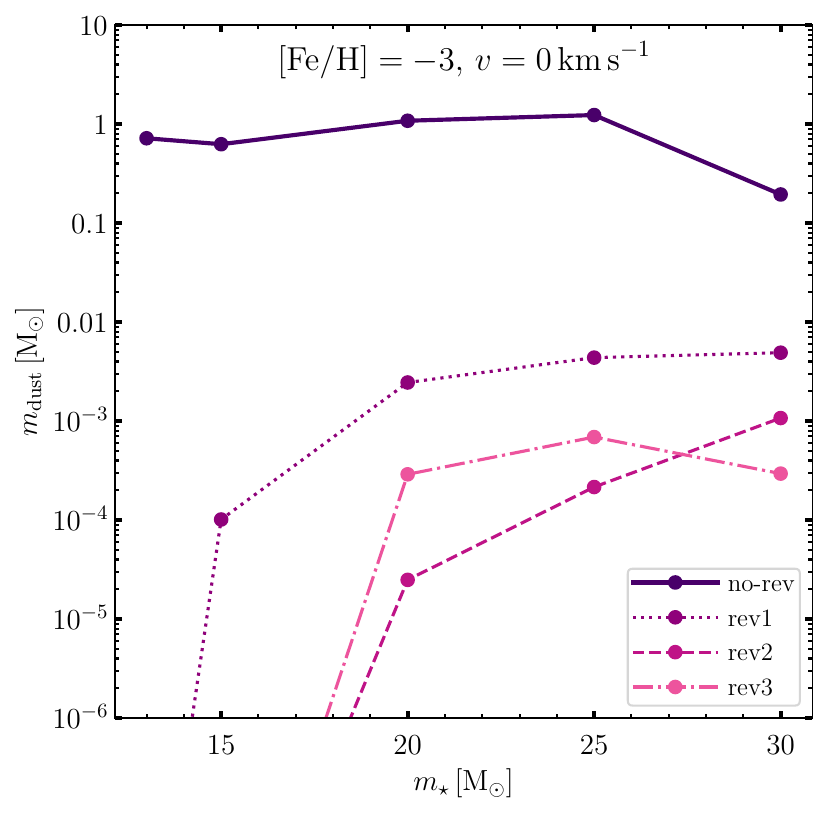}
\caption{Dust mass in SN ejecta as a function of the initial mass of non-rotating progenitor stars. Each panel corresponds to a different metallicity: $\text{[Fe/H]}=0$ (top left), $\text{[Fe/H]}=-1$ (top right), $\text{[Fe/H]}=-2$ (bottom left), and $\text{[Fe/H]}=-3$ (bottom right). The solid, dotted, dashed, and dash-dotted lines represent no-rev, rev1, rev2, and rev3 models in each panel. }
\label{fig:totaldust_nonrot}
\end{figure*}

\begin{figure*}
\includegraphics[width=.5\linewidth]{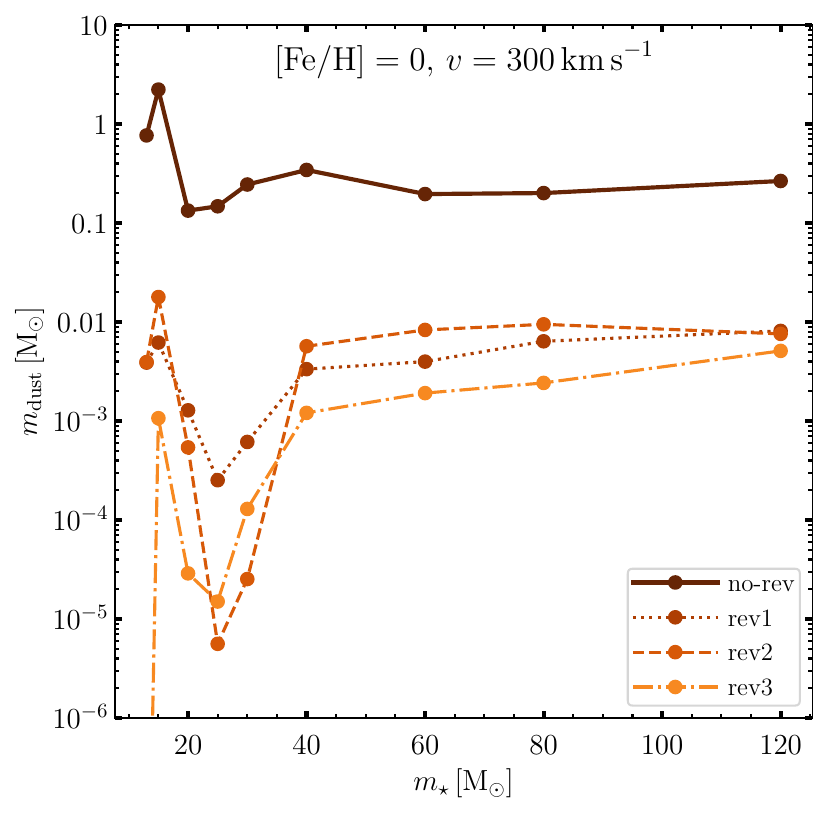}
\includegraphics[width=.5\linewidth]{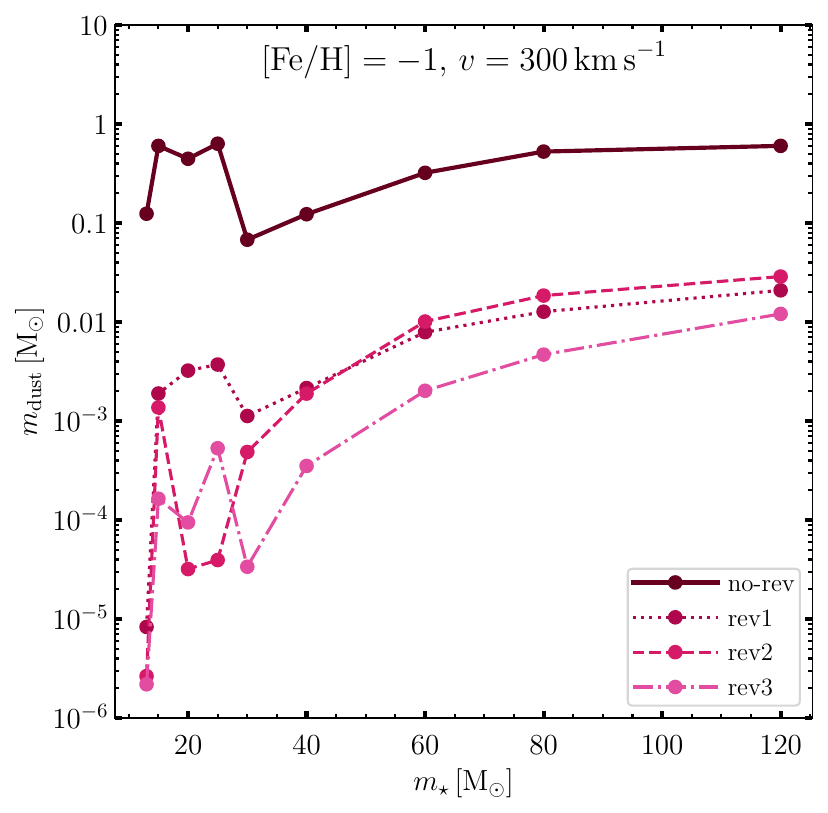}
\includegraphics[width=.5\linewidth]{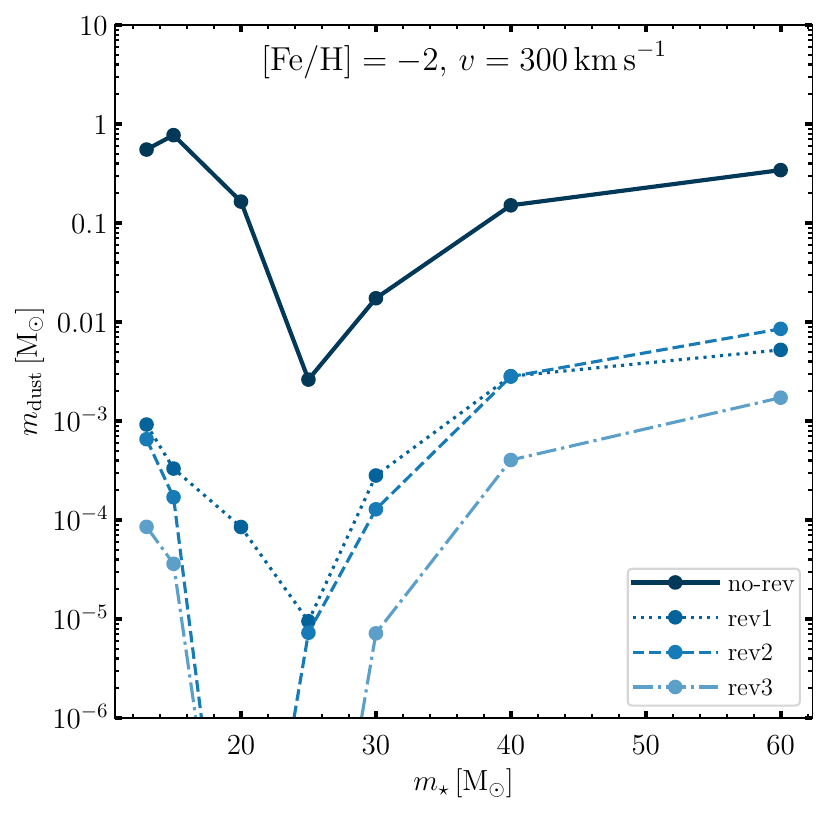}
\includegraphics[width=.5\linewidth]{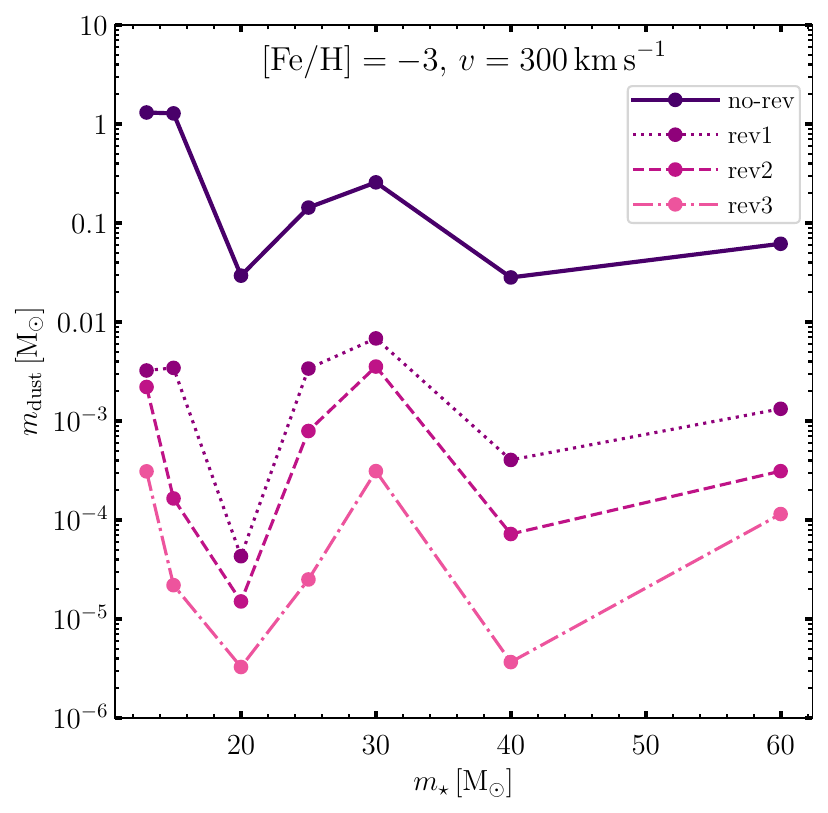}
\caption{Same as Fig. \ref{fig:totaldust_nonrot}, but for rotating progenitor stars.}
\label{fig:totaldust_rot}
\end{figure*}

We apply the \texttt{GRASHrev} model to the grids of non-rotating and rotating SN dust yields described in the previous sections. 
For each SN model, we quantify the effects of the reverse shock on the mass, composition and size distribution of the ejecta dust assuming three different values for the surrounding ISM density, $\rho_{\rm ISM} = 1.5 \times 10^{-25}, 1.5 \times 10^{-24},$ and $1.5 \times 10^{-23}\,\mathrm{g\,cm^{-3}}$, which correspond to $n_{\rm ISM} \simeq 0.05, \,0.5$ and $5\, \mathrm{cm^{-3}}$, for an assumed molecular weight of $\mu = 1.69$. The resulting dust yields are labeled as rev1, rev2 and rev3, respectively. The no-rev case indicates the results described above, before the passage of the reverse shock, but could also be indicative of the expected yields if the explosions occur in a very under-dense region of the ISM (with $n_{\rm ISM} \ll 0.05\,\mathrm{cm^{-3}}$).

The total dust mass is defined as 
\begin{equation}
    m_\mathrm{dust} = \sum_i\int_{a_\mathrm{min}}^{a_\mathrm{max}} \left(\frac{\mathrm{d}n}{\mathrm{d}a}\right)_i \frac43\pi a^3 \rho_i \, \mathrm{d}a,
\end{equation}
where $(\mathrm{d}n/\mathrm{d}a)_i$ is the dust size distribution of $i$-th grain species defined within the range of minimum size $a_\mathrm{min}$ and maximum size $a_\mathrm{max}$, and $\rho_i$ is the grain density of the $i$-th grain species. The grain densities for each specie are summarized in \cite{bocchio2016}, and the iron density is set to be $7.9\,\mathrm{g\,cm^{-3}}$ \citep{marsh1980}.

The total surviving dust mass at $10^6\,\mathrm{yr}$ after the explosion is shown as a function of the initial SN progenitor mass and metallicity for non-rotating and rotating models in Figs. \ref{fig:totaldust_nonrot} and \ref{fig:totaldust_rot}. 
In each panel, the three dashed lines indicate the results of the three reverse shock (rev shock) models. Consistent with previous studies \citep{bianchi2007, bocchio2016}, we find that the destructive effect of the reverse shock increases with the ambient interstellar medium density, $n_{\rm ISM}$, with the smallest surviving dust mass typically observed in the rev3 model. Hence, the mean environmental density where the explosions occur appears to be a critical factor to assess the efficiency of interstellar dust enrichment by SNe. 
However, the extent of destruction caused by the reverse shock strongly depends on the initial grain size distribution, and rev3 does not always yield the lowest surviving dust mass.

As a general trend, grains formed in the ejecta of the most massive SN progenitors appear to suffer less destruction compared to less massive ones, independently of the initial metallicity and rotation rate. In the cases of non-rotating progenitors with $13$ and $15\,M_\odot$, the dust grains are almost completely destroyed by the reverse shock. This implies that - although dust is generally most efficiently synthesized in the SN ejecta of $15 \text{--} 25 \, M_\odot$ stars - the explosions of more massive stars may lead to larger dust enrichment if the explosions occur in a high density region of the ISM. We find that the maximum mass of dust surviving the passage of the reverse shock is released by SN explosions with $120 \, M_\odot$ progenitors exploding in a region of the ISM with $n_\mathrm{ISM}=0.5\,\mathrm{cm^{-3}}$: for the non-rotating $\text{[Fe/H]}=0$ model, the surviving dust mass is $m_{\rm dust} \simeq 0.02 \,M_\odot$ and corresponds to $\simeq 4\%$ of the initial dust mass formed in the ejecta. A similar surviving mass fraction ($\simeq 5\%$) is found for the rotating $\text{[Fe/H]}=-1$ model, with a final dust mass of $m_{\rm dust} \simeq 0.03 \, M_\odot$.

\subsection{Results for non-rotating models}

\begin{figure*}

\centering
\includegraphics[width=0.9\linewidth]{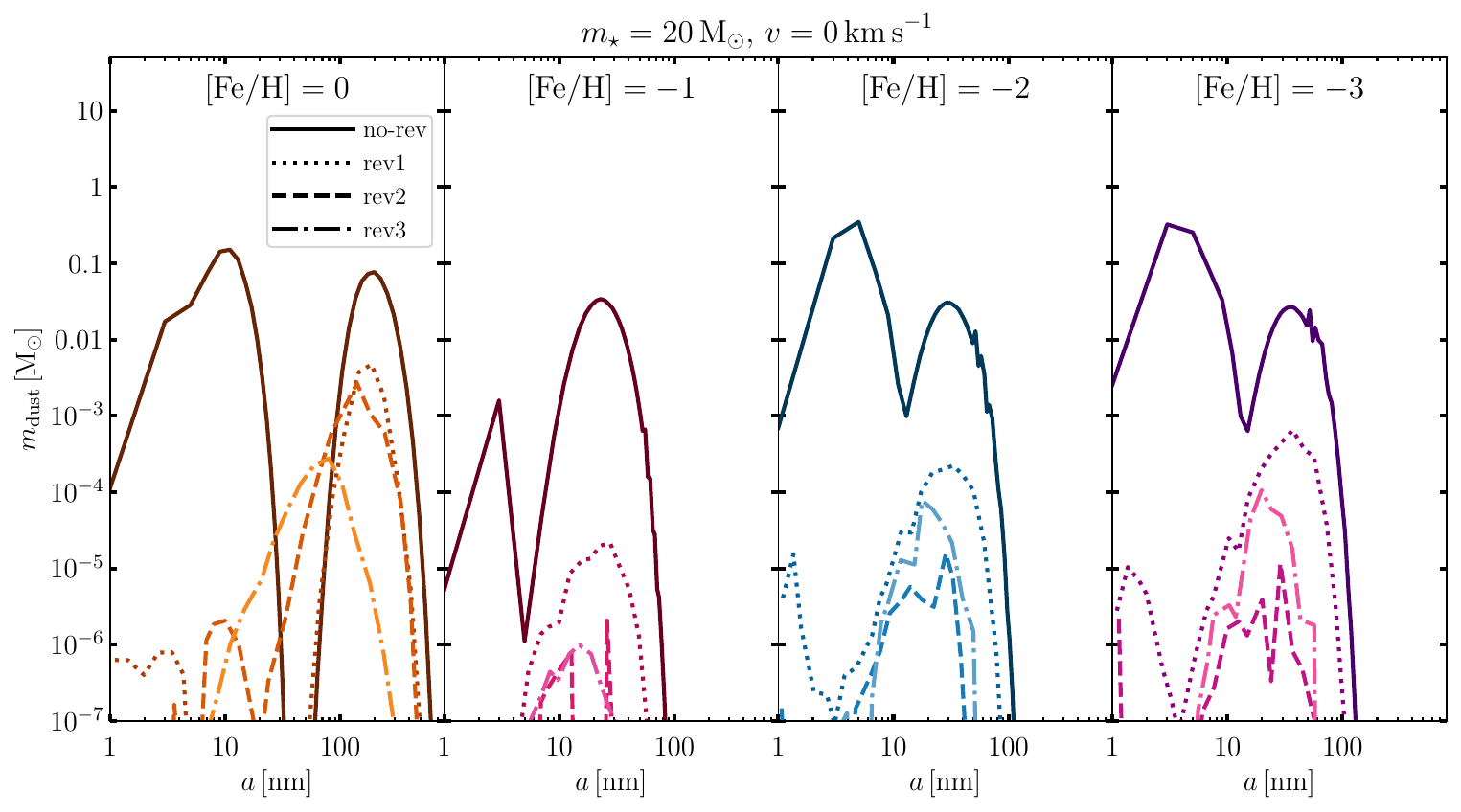}
\caption{
Grain size distribution for a non-rotating progenitor with a mass of $20 \, M_\odot$. The horizontal axis represents the grain size while the vertical axis indicates the corresponding dust mass. From left to right, the panels show results for initial stellar metallicity of $\text{[Fe/H]}=0,\,-1,\,-2,$ and $-3$, respectively. In each panel, the solid, dotted, dashed, and dash-dotted lines correspond to the no-reverse shock (no-rev), rev1, rev2, and rev3 models, respectively.}
\label{fig:distri_M20V0}
\end{figure*}

\begin{figure}
\centering
\includegraphics[width=\linewidth]{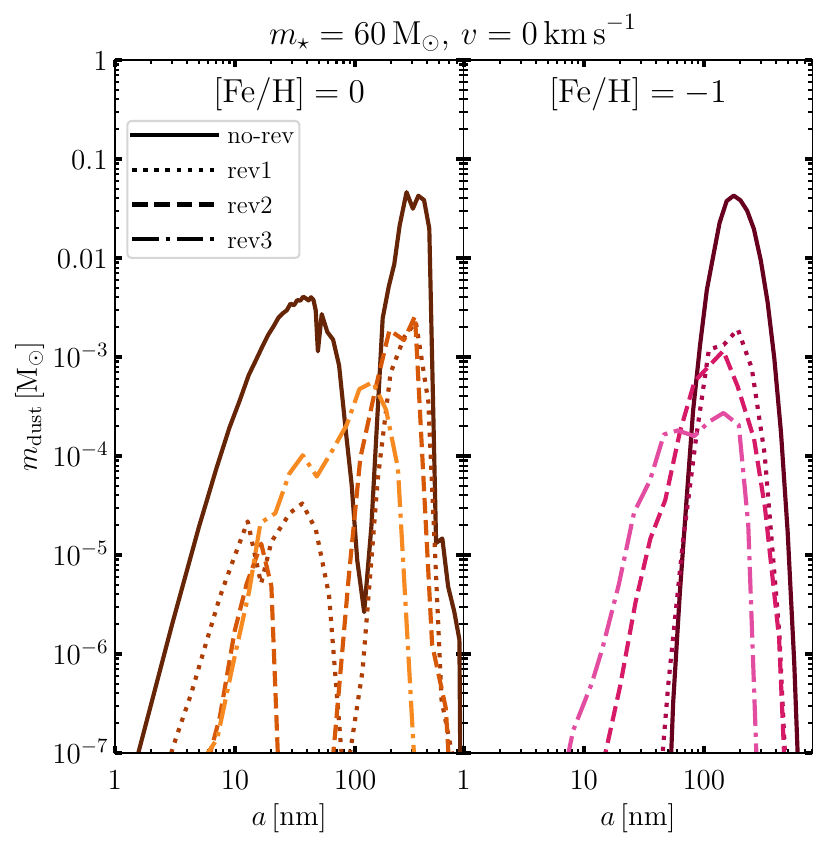}
\caption{Same as Fig. \ref{fig:distri_M20V0}, but the progenitor mass is $60\,M_\odot$.}
\label{fig:distri_M60V0}
\end{figure}

In Fig.~\ref{fig:distri_M20V0}, we present the grain size distribution for a non-rotating progenitor with a mass of $20\,M_\odot$, in order to evaluate the impact of the reverse shock. On the vertical axis, we show the dust mass contributed by each 

grain size bin $a_j$, computed as
\begin{equation}
m_\mathrm{dust}(a_j) = \sum_i \int_{a_j}^{a_j+\Delta a_j} \left(\frac{\mathrm{d}n}{\mathrm{d}a}\right)_i \frac43\pi a^3 \rho_i \, \mathrm{d}a,
\end{equation}
where $\Delta a_j$ is the size of the $j$-th bin.

From left to right, each panel corresponds to a different initial metallicity: $\text{[Fe/H]} = 0,-1,-2$, and $-3$. The solid, dotted, dashed, and dash-dotted lines represent the no-reverse shock (no-rev), rev1, rev2, and rev3 models, respectively.

In the no-rev models, the size distribution is bimodal, with peaks at $a \simeq 10\,\mathrm{nm}$ and $a \simeq 100\,\mathrm{nm}$ for $\text{[Fe/H]} = 0$. These peaks shift toward smaller sizes as the metallicity decreases. The larger grains are primarily composed of amorphous carbon (\ce{AC}), while the smaller grain population is dominated by \ce{Fe3O4} and \ce{Mg2SiO4}.

Dust grains with initial sizes $a \gtrsim 10\,\mathrm{nm}$ are more resistant to the passage of the reverse shock, with \ce{AC} contributing the majority of the surviving dust mass in most cases. 
Conversely, grains with $a \lesssim 10\,\mathrm{nm}$ are significantly destroyed by the reverse shock, despite their initially large dust mass fractions, particularly at $\text{[Fe/H]} = 0,-2$, and $-3$.

Similar results are found when analyzing other initial progenitor masses. In some cases, the distribution is not bimodal, but rather lognormal with a single broad peak. As an example, we show in Fig. \ref{fig:distri_M60V0} the grain size distribution of a $60\,M_\odot$ progenitor with $\text{[Fe/H]}=0$ (left panel) and $\text{[Fe/H]}=-1$ (right panel). The no-rev model with $\text{[Fe/H]}=0$ has a bimodal distribution separated in $a\sim 100\,\mathrm{nm}$, which is composed of \ce{Fe} grains with $a\lesssim 100\,\mathrm{nm}$ and \ce{AC} grains with $a\gtrsim100\,\mathrm{nm}$. After the passage of the reverse shock, the main contribution of the dust yield is \ce{AC}, although the \ce{Fe} grains also survive. For $\text{[Fe/H]}=-1$, the no-rev model is characterized by a log-normal distribution consisting of \ce{AC} grains. As a result, even in this case, the main contribution to the surviving dust mass is by AC grains, with a size distribution that is shifted towards smaller grains with increasing $n_\mathrm{ISM}$.

\begin{figure*}

\includegraphics[width=9.0cm]{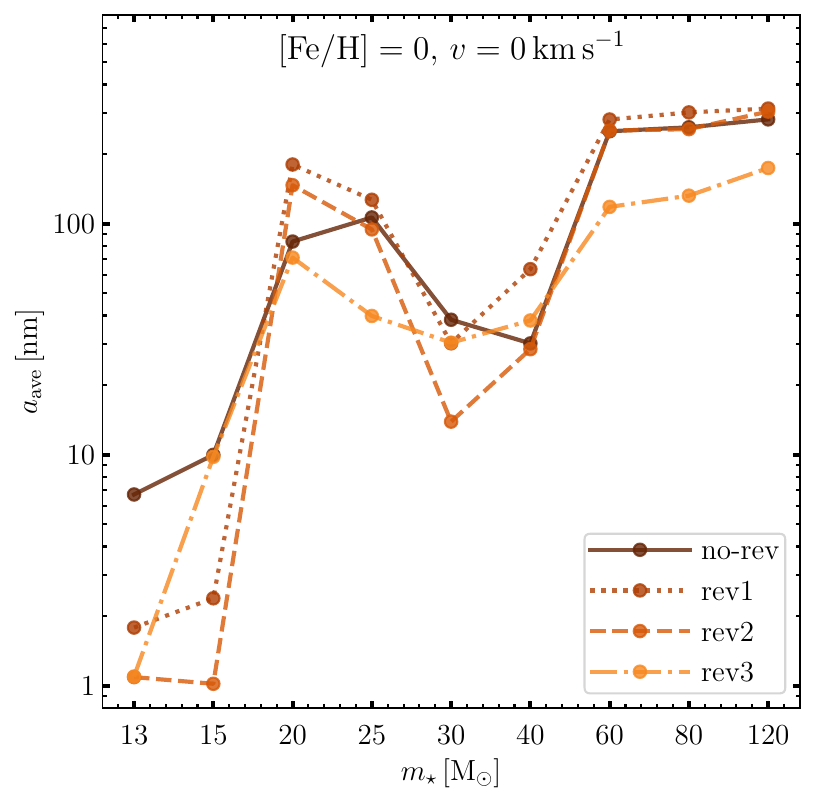}
\includegraphics[width=9.0cm]{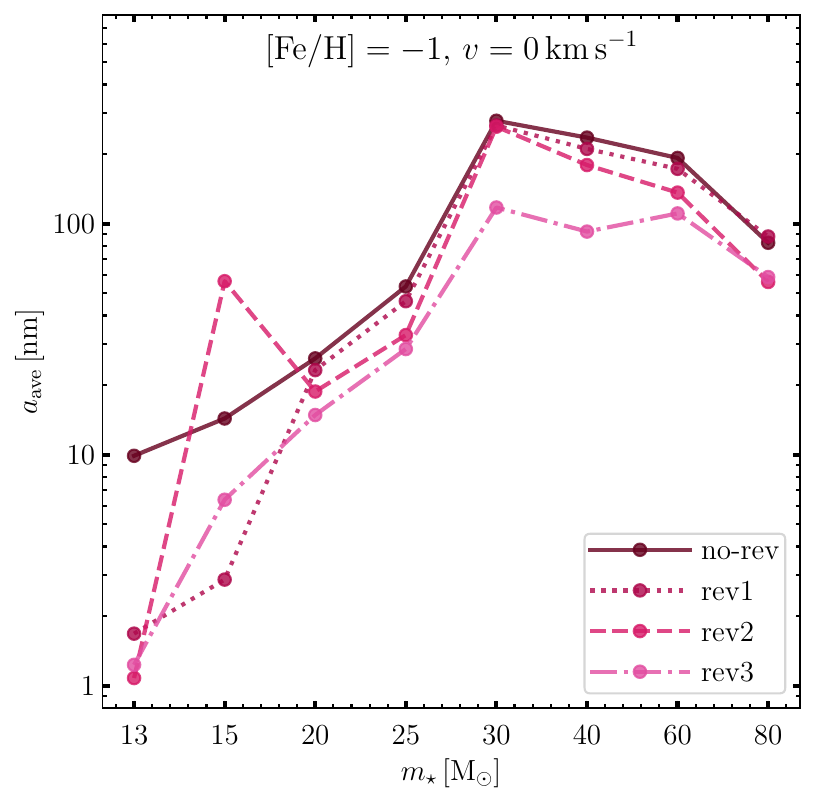}
\includegraphics[width=9.0cm]{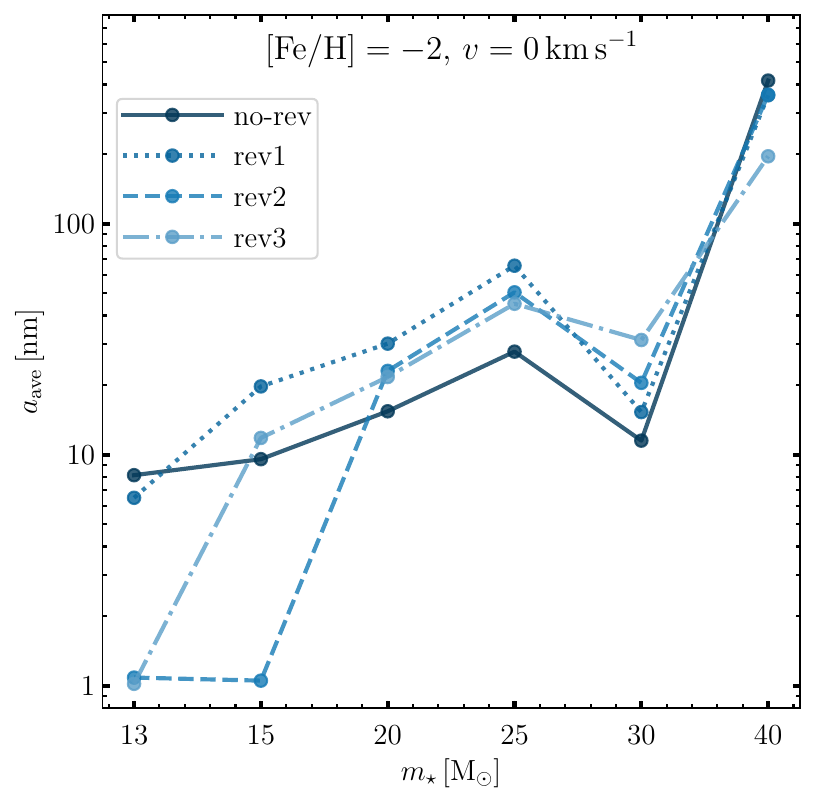}
\includegraphics[width=9.0cm]{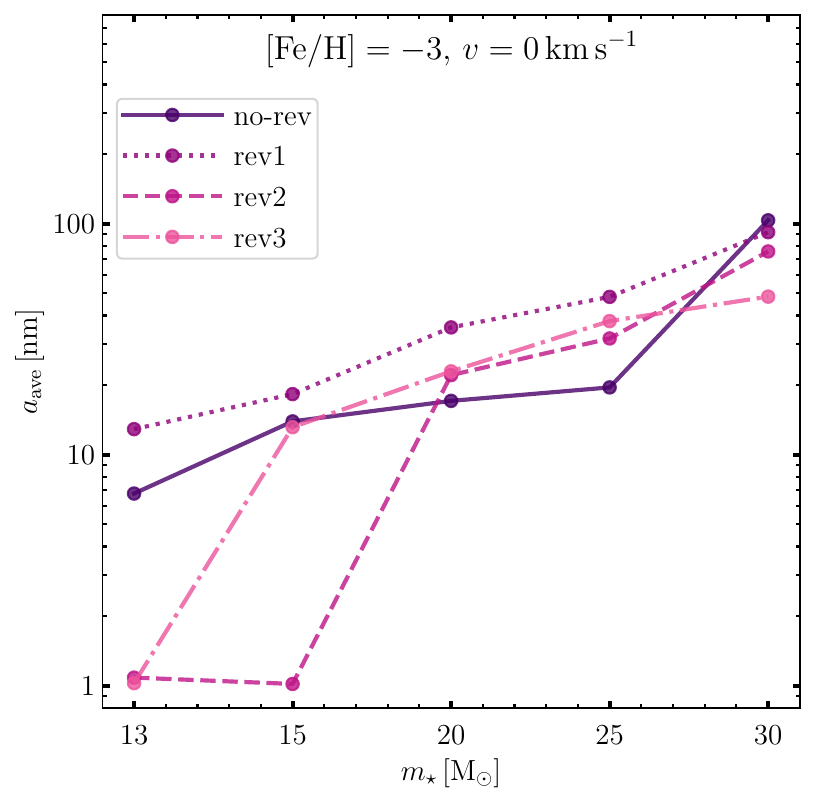}
\caption{Dust-mass-weighted average grain size as a function of the mass of the non-rotating progenitor star.}
\label{fig:average_size_v0}
\end{figure*}

Fig. \ref{fig:average_size_v0} shows the average grain size weighted by dust mass for non-rotating progenitor stars, computed as:

\begin{equation}
a_\mathrm{ave} = \frac{\sum_j a_j m_\mathrm{dust}(a_j)}{m_\mathrm{dust}},
\end{equation}

as a function of the initial stellar progenitor mass. The solid, dotted, dashed, and dash-dotted lines correspond to no-rev, rev1, rev2, and rev3 models, respectively.

Compared to the no-rev models, the reverse shock models predict smaller average grain sizes for less massive progenitor stars, such as those with $13\,M_\odot$ and $15\,M_\odot$, particularly at $\text{[Fe/H]} = 0$ and $-1$. In contrast, more massive progenitor stars tend to produce larger dust grains, with average sizes $a_{\rm ave} \gtrsim 100\,\mathrm{nm}$ after the passage of the reverse shock. The larger amorphous carbon (\ce{AC}) grains ejected in the no-rev models are more likely to survive the destructive effects of the reverse shock.

A general trend observed is that the average grain size decreases with increasing ambient interstellar medium density ($n_\mathrm{ISM}$), reflecting the enhanced destruction of smaller grains. However, the reverse shock can also significantly erode larger grains as $n_\mathrm{ISM}$ increases. For progenitors of $13\,M_\odot$ and $15\,M_\odot$, the surviving dust masses are typically $\lesssim 10^{-6}\,M_\odot$, with most of the grains reduced to sizes around $\sim 1\,\mathrm{nm}$ due to the impact of the reverse shock.

In contrast, large dust grains ($\gtrsim 100\,\mathrm{nm}$) are predominantly produced in the ejecta of more massive progenitors with $\text{[Fe/H]} = 0$, $-1$, and $-2$.

%FFFFFFFFFFFFFFFFFFFFFFFFFFFFFFFFFFFFFFFFFFFFFFFFFFFFFFFFF

%FFFFFFFFFFFFFFFFFFFFFFFFFFFFFFFFFFFFFFFFFFFFFFFFFFFFFFFFF
\begin{figure*}

\includegraphics[width=.5\linewidth]{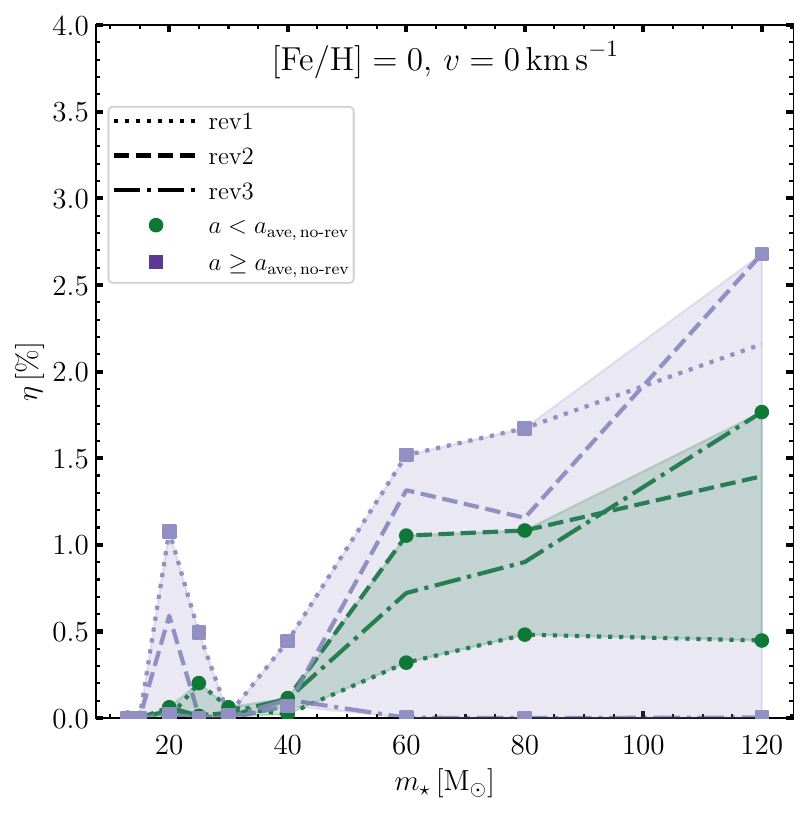}
\includegraphics[width=.5\linewidth]{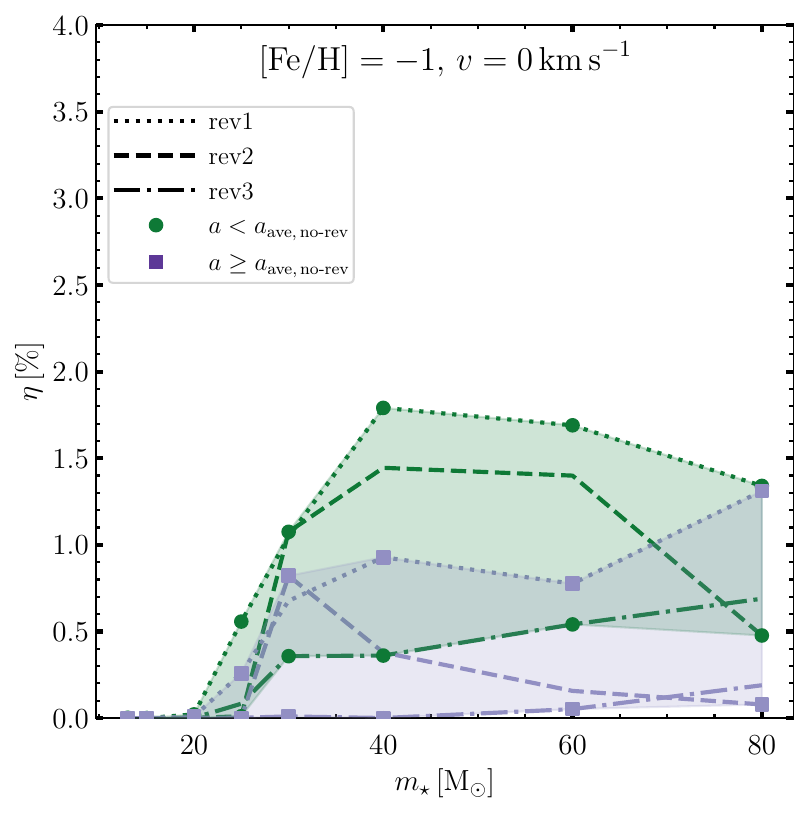}
\caption{Surviving dust mass fractions as a function of the mass of non-rotating progenitor stars. The left and right panels represent the metallicity of $\text{[Fe/H]}=0$ and $-1$, respectively. In each panel, the dotted, dashed, and dash-dotted lines correspond to rev1, rev2, and rev3 models. The green and purple colors indicate the minimum and maximum mass fractions corresponding to grain sizes with $a<a_\text{ave,\,no-rev}$ and $a\geq a_\text{ave,\,no-rev}$, respectively. $a_\text{ave,\,no-rev}$ is the dust-mass weighted average grain size for each non-rotating progenitor star in the no-rev model shown in Fig. \ref{fig:average_size_v0}. The symbols and shaded areas represent the minimum and maximum mass fractions for each progenitor mass. We shall note here that $\eta$ can be interpreted as the true surviving fraction for large grains. For smaller grains, instead, this quantity does not reflect the fraction of pre-shock small grains that survives the reverse shock; rather, it compares the post-shock mass in small grains - which also results from partial destruction of larger grains - to the initial pre-shock value (see text).
} 
\label{fig:fraction_nonrot}
\end{figure*}

In Fig.~\ref{fig:fraction_nonrot}, we show the surviving dust mass fraction, $\eta$, defined as the ratio of the dust mass after the passage of the reverse shock to the initial dust mass, plotted as a function of the initial mass of the progenitor star. The left and right panels correspond to progenitors with initial metallicity of $\text{[Fe/H]} = 0$ and $-1$, respectively.

We divide the surviving dust fraction into two categories: grains with sizes $a < a_\text{ave,no-rev}$ (green circles) and $a \geq a_\text{ave,no-rev}$ (purple squares), where $a_\text{ave,no-rev}$ represents the dust-mass-weighted average grain size for each non-rotating progenitor in the no-rev model (shown as solid lines in Fig.~\ref{fig:average_size_v0}). The dotted, dashed, and dash-dotted lines correspond to the rev1, rev2, and rev3 models, respectively. Shaded areas indicate the range between the maximum and minimum surviving fractions across different ISM densities.

For $\text{[Fe/H]} = 0$, the surviving mass fraction of larger grains 
is higher than that of smaller grains in massive progenitor stars, as large amorphous carbon (\ce{AC}) grains are more likely to survive the reverse shock. The maximum surviving fraction for large grains is 2.7\% in the rev2 model for the $120\,M_\odot$ progenitor. For small grains, the highest surviving fraction is 1.8\% in the rev3 model for the same progenitor.

In contrast, for $\text{[Fe/H]} = -1$, the mass fraction of smaller grains exceed those of larger grains for progenitors with $40\,M_\odot$ and $60\,M_\odot$. We shall note here that this quantity does not reflect the fraction of pre-shock small grains that survives the reverse shock; rather, it compares the post-shock mass in small grains - which also results from partial destruction of larger grains - to the initial pre-shock value. 

As shown in Fig.~\ref{fig:distri_M60V0}, the grain size distribution at this metallicity is not bimodal, and the dust mass below and above the average grain size is roughly comparable. In these cases, the reverse shock fragments initially large grains into smaller ones, resulting into a higher fraction of small grains. For all other progenitor masses—except $13\,M_\odot$ and $15\,M_\odot$—the size distributions are bimodal, and the surviving fractions for grains with $a \geq a_\text{ave,no-rev}$ exceed those for smaller grains.

At $\text{[Fe/H]} = -1$, the maximum surviving fraction for large grains is 1.3\%, observed in the rev1 model for the $80\,M_\odot$ progenitor, while the maximum for small grains is 1.8\%, also occurring in the rev1 model for the $40\,M_\odot$ progenitor.

\subsection{Results for rotating models}

%FFFFFFFFFFFFFFFFFFFFFFFFFFFFFFFFFFFFFFFFFFFFFFFFFFFFFFFFF

\begin{figure*}
\centering
\includegraphics[width=.9\linewidth]{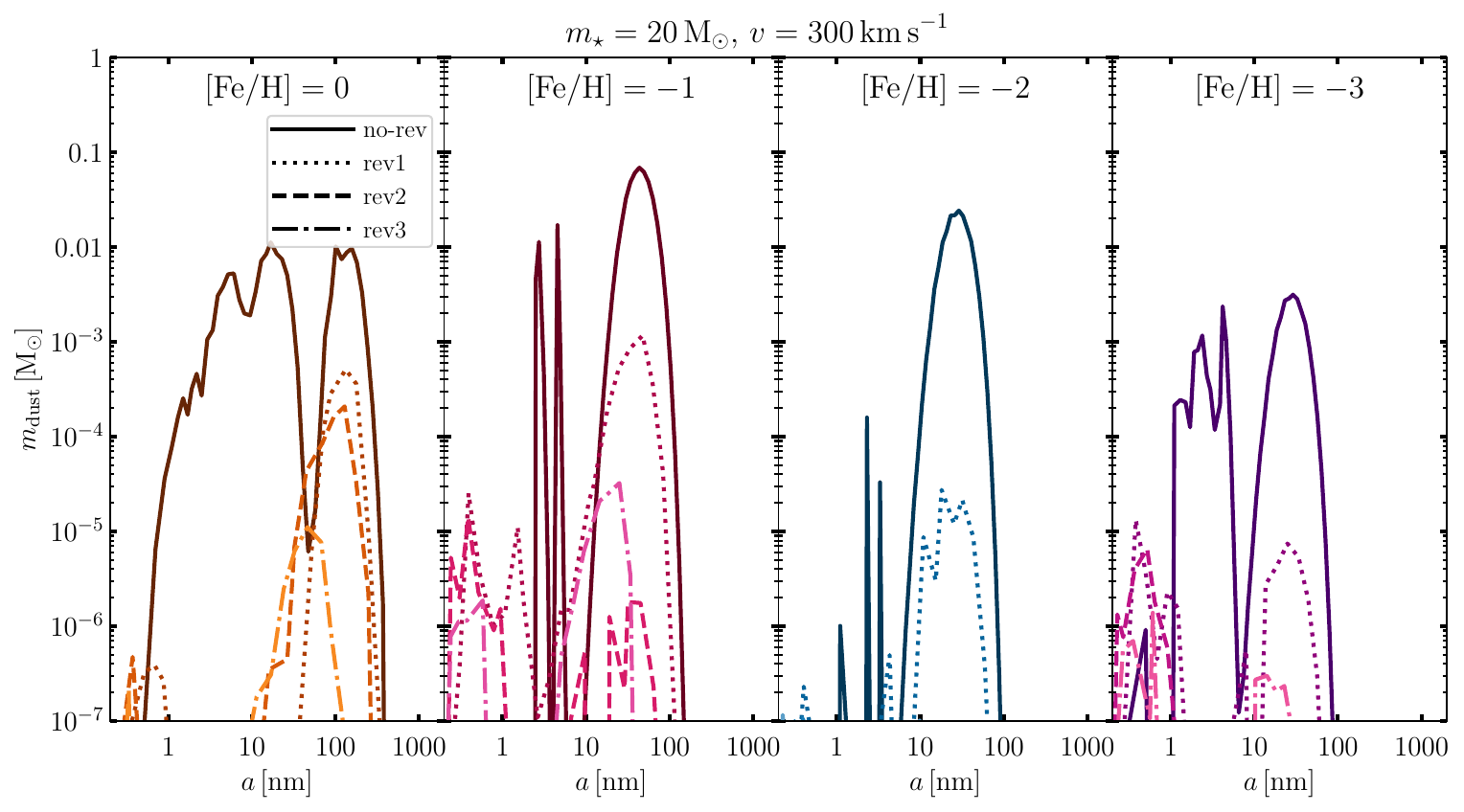}
\caption{Grain size distribution for a rotating progenitor with a mass of $20 \, M_\odot$. From left to right, the panels show results for initial stellar metallicity of $\text{[Fe/H]}=0,\,-1,\,-2,$ and $-3$, respectively. In each panel, the solid, dotted, dashed, and dash-dotted lines correspond to the no-reverse shock (no-rev), rev1, rev2, and rev3 models, respectively.}
\label{fig:distri_M20V300}
\end{figure*}

Figure~\ref{fig:distri_M20V300} shows the grain size distribution for rotating progenitor stars with a mass of $20\,M_\odot$. Similar to the non-rotating models, larger grains are more likely to survive the passage of the reverse shock. In the no-rev model, amorphous carbon (\ce{AC}) grains dominate the larger grain population. In contrast, smaller grains are almost entirely destroyed by the reverse shock at $\text{[Fe/H]} = 0$.

For the rotating progenitor with $\text{[Fe/H]} = -2$, no dust grains are ejected after the reverse shock in environments with interstellar medium (ISM) densities of $n_\mathrm{ISM} = 0.5$ and $5\,\mathrm{cm^{-3}}$, indicating complete destruction. However, for metallicity of $\text{[Fe/H]} = -1$ and $-3$, small \ce{MgSiO3} and \ce{Mg2SiO4} grains with sizes around $\sim 1,\mathrm{nm}$ survive in lower-density ISM conditions ($n_\mathrm{ISM} = 0.05$ and $0.5\,\mathrm{cm^{-3}}$).

Among the reverse shock models, the rev1 model shows the highest survival of large grains after the shock, compared to rev2 and rev3.

\begin{figure*}

\includegraphics[width=.5\linewidth]{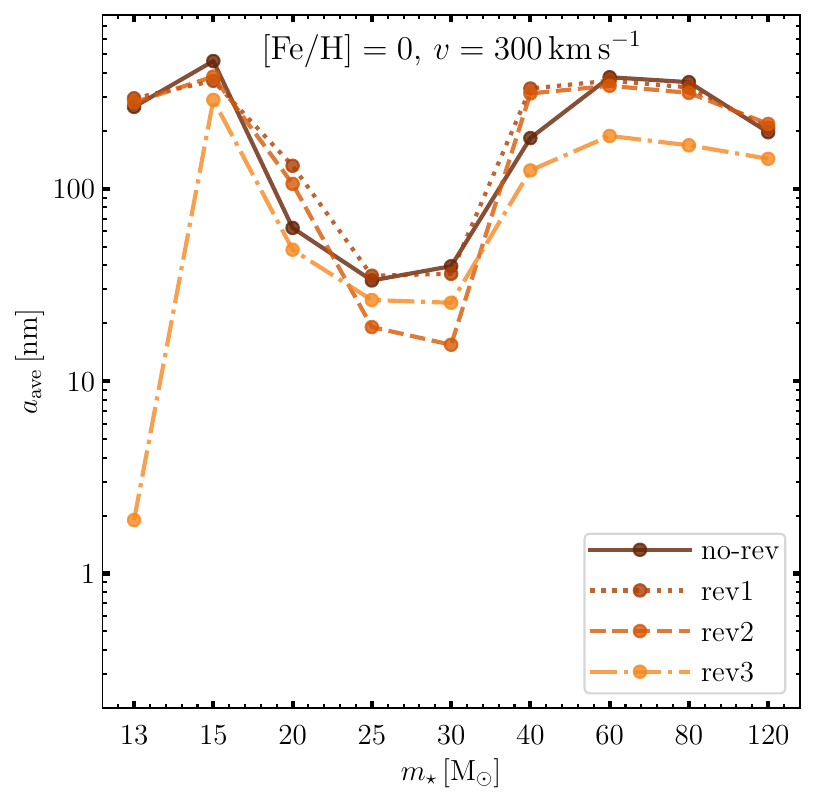}
\includegraphics[width=.5\linewidth]{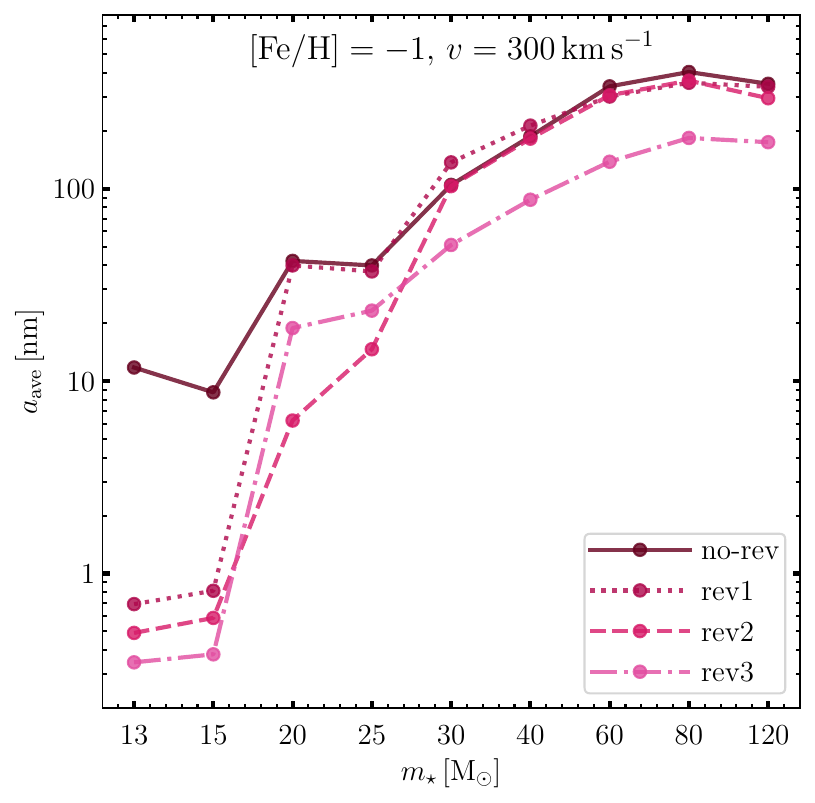}
\includegraphics[width=.5\linewidth]{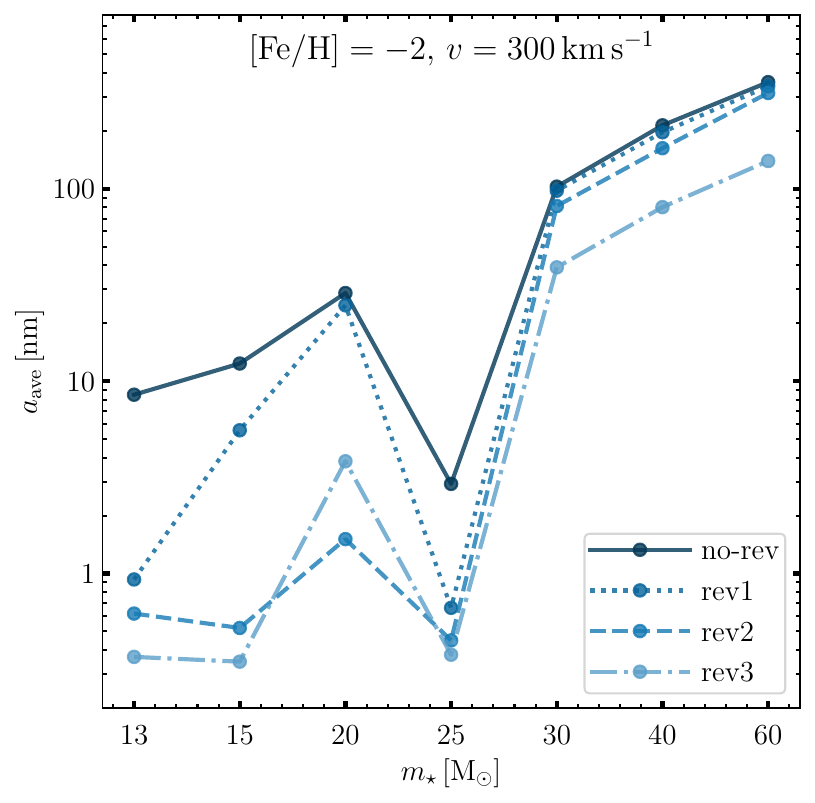}
\includegraphics[width=.5\linewidth]{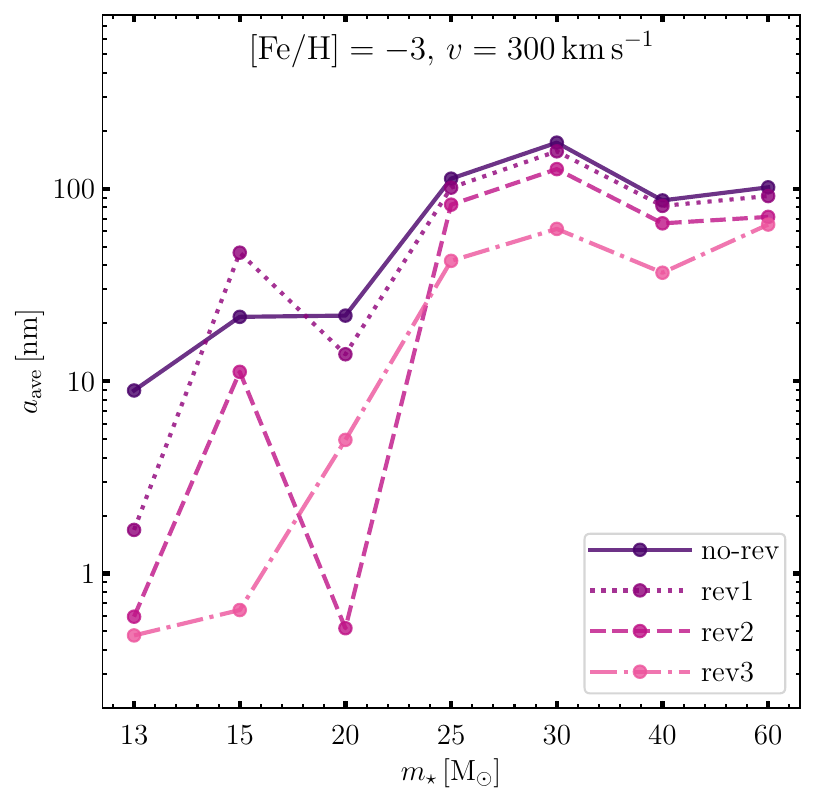}

\caption{Same as Fig. \ref{fig:average_size_v0}, but for rotating progenitor stars.}
\label{fig:average_size_v300}
\end{figure*}

In Fig.~\ref{fig:average_size_v300}, we present the average grain size - weighted by dust mass - for rotating progenitor stars. Compared to their non-rotating counterparts, the rotating models generally produce larger average grain sizes. This is particularly evident for massive progenitors, which eject grains with $a_\mathrm{ave} \gtrsim 100\,\mathrm{nm}$ even after the passage of the reverse shock.

In contrast, lower-mass progenitors such as those with $13\,M_\odot$ and $15\,M_\odot$ produce significantly smaller grains, especially at metallicity of $\text{[Fe/H]} = -1$, $-2$, and $-3$. The reverse shock becomes increasingly effective at reducing the size of large grains as the ambient ISM density increases, highlighting the strong dependence of dust survival on both stellar mass and environmental conditions.

%FFFFFFFFFFFFFFFFFFFFFFFFFFFFFFFFFFFFFFFFFFFFFFFFFFFFFFFFF
\begin{figure*}
\includegraphics[width=9.0cm]{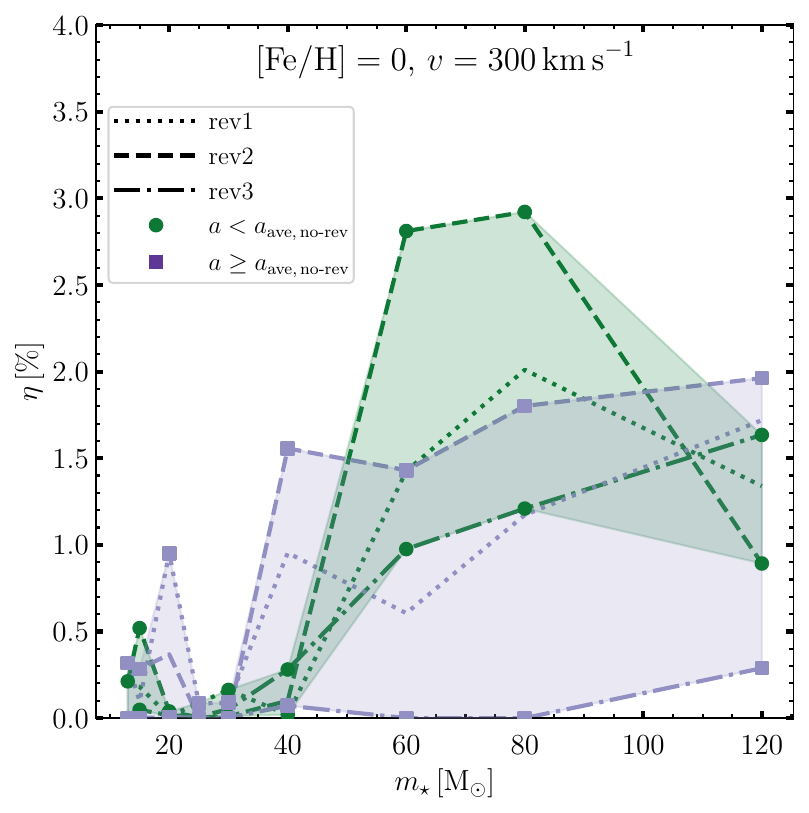}
\includegraphics[width=9.0cm]{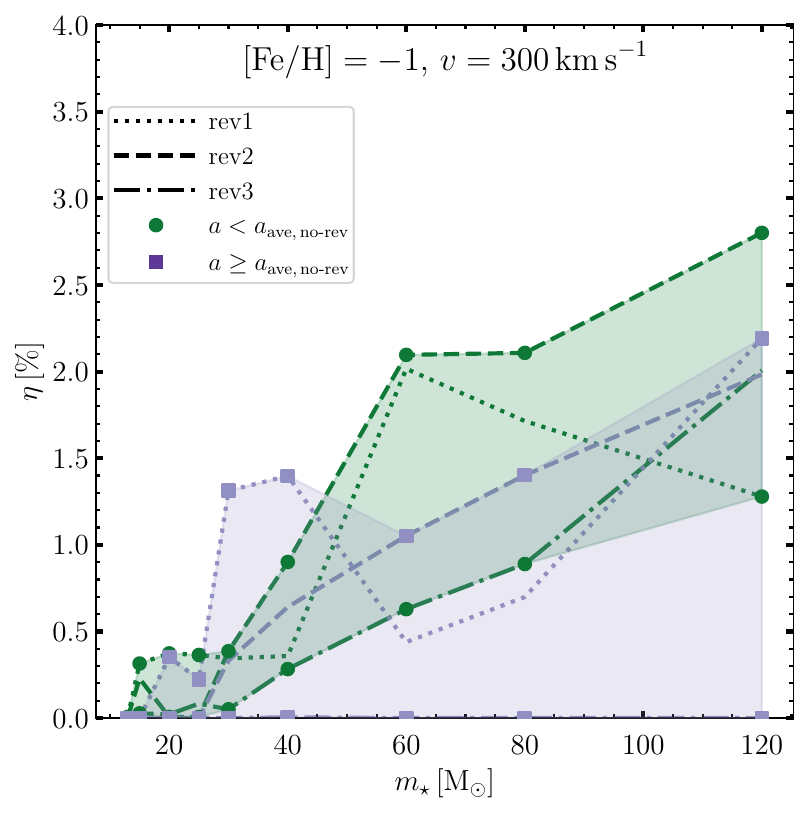}
\caption{Same as Fig. \ref{fig:fraction_nonrot}, but for rotating models.}
\label{fig:fraction_rot}
\end{figure*}
%FFFFFFFFFFFFFFFFFFFFFFFFFFFFFFFFFFFFFFFFFFFFFFFFFFFFFFFFF

Figure~\ref{fig:fraction_rot} shows the surviving dust mass fractions for the rotating progenitor models. Compared to the non-rotating cases, the surviving fractions of small grains are systematically higher for massive progenitors.

For both metallicity, $\text{[Fe/H]} = 0$ and $-1$, smaller grains exhibit higher surviving fractions than larger grains. This trend arises because the reverse shock tends to fragment larger grains, increasing the population of smaller grains. Notably, the $120\,M_\odot$ progenitor at $\text{[Fe/H]} = 0$ shows a particularly high surviving fraction of small grains in both the rev1 and rev2 models. Specifically, at $\text{[Fe/H]} = 0$, the maximum surviving fraction of large grains is 2.0\% in the rev2 model for the $120\,M_\odot$ progenitor, while the highest surviving fraction of small grains reaches 2.9\% in the rev2 model for the $80\,M_\odot$ progenitor.

At $\text{[Fe/H]} = -1$, the surviving fractions for both small and large grains remain relatively high. For the $120\,M_\odot$ progenitor, the surviving fraction of small grains reaches at least 2.8\% in the rev2 model, while large grains survive at a rate of 2.2\% for the same progenitor in the rev2 model.

\section{Discussion}\label{sec:discussion}

The results presented in the previous sections provide a quantitative estimate of the effective dust yields produced by massive stars, based on a homogeneous grid of supernova (SN) models spanning a range of initial stellar masses ($13\text{--}120\,M_{\odot}$), two initial rotation velocities ($v = 0$ and $v = 300\,\mathrm{km\,s^{-1}}$), and four different initial metallicity: $\text{[Fe/H]} = 0$, $-1$, $-2$, and $-3$ \citep{limongi2018}. A key finding of our study is that only a small fraction - typically less than 10\% - of the initially formed dust survives the reverse shock and is subsequently injected into the interstellar medium (ISM). However, the exact survival yield depends sensitively on the progenitor's mass, metallicity, rotation rate, and the local ISM density at the time of the explosion.

It is important to acknowledge the limitations of our approach. 

The initial grain populations used in our calculations are based on the predictions of \citet{marassi2019}, which apply classical nucleation theory within a homogeneous, spherically symmetric, uniformly mixed ejecta model. This assumption significantly simplifies the complex and inherently asymmetric nature of real SN explosions \citep{bruenn2016}. In reality, dust nucleation likely occurs in clumpy ejecta, potentially leading to the formation of larger grains that are more resistant to destruction by the reverse shock (see \citealt{micelotta2018} and references therein). Conversely, clumpy environments can also enhance grain-grain collisions, promoting the fragmentation of grains into smaller sizes \citep{kirchschlager2019, kirchschlager2020}.

Moreover, we do not consider the possible reformation of dust in cold, dense shell, between the forward and reverse shocks, as suggested by recent JWST observations of stripped-envelope SN explosions interacting with their dense, hydrogen rich, circumstellar medium \citep{shahbandeh2024, tinyanont2025}. Such secondary dust formation could contribute significantly to the overall dust budget and increase the effective mass of dust that ultimately enriches the ISM.

Despite these limitations, our findings are broadly consistent with previous studies, as discussed in Section~\ref{sec:previous}, reinforcing the emerging consensus that reverse shock destruction plays a major role in regulating the efficiency of dust enrichment by core-collapse supernovae.

In our study, we have assumed that all supernovae (SNe) explode with a fixed-energy of $1.2 \times 10^{51}\,\mathrm{ergs}$. However, observations indicate that SNe span a broad range of explosion energies, from faint explosions to hypernovae \citep{nomoto2013}. In \citet{marassi2019}, an independent grid of SN models was developed and calibrated to reproduce the observed relationship between the mass of synthesized $^{56}$Ni and the progenitor mass, as shown in their Fig.~1. These Calibrated Energy (CE) models result in more massive and metal-rich ejecta, leading to more efficient dust production. In these models, the dust mass increases with progenitor mass and is dominated by silicate grains across all metallicity \citep{marassi2019}.

The choice of explosion energy calibration is clearly critical, as it directly influences the location of the mass cut, and consequently, the mass and metal composition of the ejecta. \citet{limongi2018} explored the impact of such calibration choices on metal yields and proposed a recommended set of models. In their framework, all stars with initial masses in the range $13 \text{--} 25\,M_\odot$ are assumed to produce a fixed amount of $^{56}$Ni ($0.07\,M_\odot$), while stars with initial masses greater than $25\,M_\odot$ are assumed to fail in exploding and collapse directly into black holes.

Adopting this framework and restricting dust production to progenitors with $m_\star \leq 25\,M_\odot$ would significantly reduce the total SN dust yield compared to the results obtained using our fixed-energy SN grid, particularly for rotating progenitors. 

It is important to stress that any robust estimate of the SN contribution to dust enrichment must account for the stellar initial mass function and the star formation history (see the discussion in \citealt{schneider2024}). We will further explore these aspects in the context of early dust production at high redshift in a companion paper (Otaki et al. 2025b).

Despite the large destructive effect of the reverse shock, our results show that - on very short timescales - SNe can seed the interstellar medium with carbonaceous grains with sizes ranging from $\sim 1\,\mathrm{nm}$ to $\sim 100 \, \mathrm{nm}$ ($\sim 1 \mu\mathrm{m}$) for rotating (non-rotating) models.

This is particularly interesting in light of the recent observation of the UV extinction bump at 2175 \AA \, in the galaxy JADES-GS-z6-0 at $z = 6.71$ \citep{markov2023, witstok2023b, markov2025}. This bump has been generally interpreted as due to carbon nanoparticles, although 
\citet{li2024} shows that the extinction bump arising from those grains is too broad and peaks at wavelengths that are too long to agree with what is seen in JADES-GS-z6-0. Conversely, \citet{lin2025} demonstrates that the combined electronic absorption spectra for a number of polycyclic
aromatic hydrocarbon (PAH) molecules closely reproduce the observed peak wavelength and width. Indeed, small amorphous carbon grains, similar to what we predict to form in SN ejecta, once injected into the interstellar medium, could have reacted with hydrogen and shattered, producing PAHs.

\section{Conclusions}
\label{sec:conclusions}

We have analyzed SN dust yields and grain size distributions including the destructive effect of the reverse shock generated by the interaction of the forward shock and the surrounding ISM using the \texttt{GRASHrev} model \citep{bocchio2016}.
A homogeneous grid of SN models has been considered, with initial stellar masses $13\text{--}120\,M_\odot$, two initial rotation velocities $v=0$ and $300\,\mathrm{km\,s^{-1}}$, four different initial metallicity $[\text{Fe/H}]=0,\,-1,\,-2$, and $-3$ \citep{limongi2018}, assuming that the explosions have occurred in a uniform medium with densities $n_\mathrm{ISM}\simeq 0.05,\,0.5$ and $5\,\mathrm{cm^{-3}}$.
The results are summarized as follows:

\begin{itemize}
    \item Before the reverse shock, the most efficient dust producers in non-rotating models are progenitors with masses $m_\star \sim 20 \text{--} 25\,M_\odot$, producing up to $\sim 1.2\,M_\odot$ of dust independently of their initial metallicity. For non-rotating progenitors with low metallicity $[\ce{Fe/H}]=-3$ and large stellar masses $m_\star > 30\,M_\odot$, no dust is formed in SN ejecta due to the effect of fallback. 
 
    On the other hand, the effect of rotation-induced mixing leads to more metal-enriched ejecta, and up $2.2\,M_\odot$ of dust is produced by a $m_\star = 15\,M_\odot$, $[\ce{Fe/H}]=0$ rotating progenitor. 
    
    The grain size distribution is independent of the initial rotation velocity but it is very sensitive to other progenitor properties. In general, the size distribution has a log-normal or a bimodal shape with sizes ranging from $\sim 1\,\mathrm{nm}$ to $\sim 100 \, \mathrm{nm}$ ($\sim 1 \mu\mathrm{m}$) for rotating (non-rotating) models.

    \item Depending on the initial grain size distribution, the reverse shock can significantly reduce the dust mass.
    The surviving fraction of the initial dust mass ranges from 0 (when the surviving dust mass is $< 10^{-6} M_\odot$ we effectively assume it to be zero) to a maximum of $5\%$.
    The grains formed in SN ejecta with massive stellar progenitors tend to be more resistant to destruction compared to less massive ones, independently of the initial metallicity and rotation velocity. 
    Among non-rotating models, the maximum mass of dust surviving the passage of the reverse shock is $0.02\,M_\odot$ for the progenitor with $m_\star = 120\,M_\odot$ and $[\ce{Fe/H}]=0$ exploding in a region of the ISM with $n_\mathrm{ISM} = 0.5\,\mathrm{cm^{-3}}$. 
    For the rotating models, a maximum dust mass of $0.03 M_\odot$ is released by the same progenitor model and assuming the same ISM density. 
    
    \item The grain size distribution and composition are modified by the reverse shock. 
    As a general trend, for a given progenitor, the average grain size decreases with increasing ISM density due to the enhanced destruction of small grains. The resulting size distributions are very irregular and the effect of the reverse shock is different depending on whether the initial size distribution has a bimodal or a log-normal shape. 
    We find that the average grain sizes increase with progenitor masses, ranging from 1 nm to 400 nm, with a non-monotonic behavior. These sizes are typically smaller for lower initial metallicity models. 
     Rotating progenitors produce larger grains compared to non-rotating ones.

    \item In most cases, \ce{AC} grains provide the dominant contribution to the surviving dust mass. This is because 
    in most models AC grains have initially the largest sizes and are therefore more resistant to destruction by the reverse shock while small grains composed of other species are almost completely destroyed.

\end{itemize}

Our results indicate that the contribution of SNe to early dust enrichment significantly depends on the destruction effect of the SN reverse shock and explore for the first time the impact of the initial rotation velocity of the progenitor stars. This study provides a homogeneous set of effective dust yields that can be used to explore early dust enrichment in very high redshift galaxies and their associated extinction properties. It also shows that very small carbonaceous grains can be quickly formed and released in the ISM of high redshift galaxies, in agreement with recent JWST findings \citep{markov2023, witstok2023b, markov2025}.

All tables are available in electronic form at the CDS.
The dust yield data calculated in this paper can be found at \url{https://github.com/K-Otaki/effective-supernova-dust-yields}.

\begin{acknowledgements}
    On May 7, 2023, our dear friend and colleague, Dr. Stefania Marassi, passed away. She made a fundamental contribution to this work by computing all the supernova dust yields adopted in the present analysis. This paper is dedicated to her memory.

KO, RS, and LG acknowledge support from the PRIN 2022 MUR project 2022CB3PJ3—First Light And Galaxy aSsembly (FLAGS) funded by the European Union—Next Generation EU, and from EU-Recovery Fund PNRR - National Centre for HPC, Big Data and Quantum Computing. LG acknowledges support from the Amaldi Research Center funded by the MIUR program “Dipartimento di Eccellenza” (CUP:B81I18001170001).
\end{acknowledgements}

\bibliographystyle{aa} % style aa.bst
\bibliography{reverse_shock} % your references Yourfile.bib
\begin{appendix}
\label{sec:appendix}
\onecolumn
\section{Dust Yields} \label{Appendix}

\begin{table*}[!htbp]
\centering
\caption{Initial dust yields for non-rotating models. From top to bottom the initial stellar metallicity is $\text{[Fe/H]}=0,\,-1,\,-2,\,-3$. All yields smaller than $10^{-6} M_\odot$ have been set to zero.}
\label{tab:sndust_initial_non_rotating}
\begin{tabular}{cllllllll} %modello rotante a metallicità solare massa iniziale tutte densità
$m_\star\,[M_\odot]$	&	$m_\mathrm{dust}\,[M_\odot]$ 	&	$\ce{Al2O3}\,[M_\odot]$	&	$\ce{Fe}\,[M_\odot]$	&	$\ce{Fe3O4}\,[M_\odot]$	&	$\ce{MgSiO3}\,[M_\odot]$	&	$\ce{Mg2SiO4}\,[M_\odot]$	&	$\ce{AC}\,[M_\odot]$	&	$\ce{SiO2}\,[M_\odot]$	\\
\hline
\hline
13	&0.27	&1.6$\times 10^{-4}$	&0	&0.16	&1.9$\times 10^{-6}$	&0	&0.11	&0	\\
15	&0.41	&3.2$\times 10^{-3}$	&0	&0.18	&0.010	&2.7$\times 10^{-3}$	&0.22	&1.9$\times 10^{-5}$	\\
20	&1.0	&0.027	&0	&0.43	&0	&0.16	&0.40	&0	\\
25	&1.2	&0.035	&0	&0.38	&0	&0.51	&0.081	&0.19	\\
30	&0.16	&0.011	&0	&0.041	&0	&0.11	&2.2$\times 10^{-3}$	&0	\\
40	&0.24	&0.016	&0	&0.054	&0	&0.14	&0.035	&0	\\
60	&0.28	&2.3$\times 10^{-6}$	&0.060	&0	&0	&0	&0.22	&0	\\
80	&0.34	&0	&0.078	&0	&0	&0	&0.26	&0	\\
120	&0.46	&0	&0.12	&0	&0	&0	&0.34	&0	\\
\hline
\hline
13	&0.11	&0	&0	&0	&4.2$\times 10^{-6}$	&0	&0.11	&0	\\
15	&0.19	&0	&0	&0	&1.9$\times 10^{-5}$	&2.9$\times 10^{-5}$	&0.19	&1.7$\times 10^{-6}$	\\
20	&0.32	&0	&0	&0	&1.5$\times 10^{-3}$	&8.9$\times 10^{-5}$	&0.32	&5.0$\times 10^{-6}$	\\
25	&0.34	&0	&0	&0	&5.3$\times 10^{-3}$	&1.5$\times 10^{-3}$	&0.33	&7.1$\times 10^{-6}$	\\
30	&0.21	&7.9$\times 10^{-4}$	&0	&3.2$\times 10^{-3}$	&3.2$\times 10^{-4}$	&0.013	&0.19	&1.7$\times 10^{-3}$	\\
40	&0.29	&0	&0	&0	&0	&0	&0.29	&0	\\
60	&0.22	&0	&0	&0	&0	&0	&0.22	&0	\\
80	&0.18	&1.4$\times 10^{-3}$	&0	&5.5$\times 10^{-3}$	&6.9$\times 10^{-4}$	&0.025	&0.14	&3.0$\times 10^{-3}$	\\
\hline
\hline
13	&0.54	&0	&0	&0.37	&2.2$\times 10^{-6}$	&0	&0.17	&0	\\
15	&0.85	&1.4$\times 10^{-4}$	&0	&0.59	&1.1$\times 10^{-3}$	&3.5$\times 10^{-4}$	&0.26	&3.1$\times 10^{-6}$	\\
20	&1.0	&4.4$\times 10^{-3}$	&0	&0.27	&6.6$\times 10^{-3}$	&0.35	&0.37	&0.028	\\
25	&1.2	&3.8$\times 10^{-3}$	&0	&0.36	&0.014	&0.34	&0.30	&0.15	\\
30	&0.014	&1.6$\times 10^{-3}$	&0	&5.5$\times 10^{-4}$	&0	&8.1$\times 10^{-3}$	&3.5$\times 10^{-3}$	&0	\\
40	&0.021	&0	&0	&0	&0	&0	&0.021	&0	\\
\hline
\hline
13	&0.72	&3.9$\times 10^{-5}$	&0	&0.55	&0	&0	&0.17	&0	\\
15	&0.63	&0	&0	&0.25	&0.090	&3.7$\times 10^{-3}$	&0.28	&2.5$\times 10^{-4}$	\\
20	&1.1	&2.5$\times 10^{-3}$	&0	&0.34	&7.4$\times 10^{-3}$	&0.30	&0.38	&0.050	\\
25	&1.2	&6.7$\times 10^{-3}$	&0	&0.40	&0.031	&0.35	&0.26	&0.19	\\
30	&0.19	&1.1$\times 10^{-3}$	&0	&0	&9.7$\times 10^{-4}$	&8.1$\times 10^{-3}$	&0.18	&0	\\
\hline
\hline
\end{tabular}
\end{table*}

\begin{table*}[!htbp]
\centering
\caption{Initial dust yields for rotating models. From top to bottom the initial stellar metallicity is $\text{[Fe/H]} = 0,\, -1,\,-2,\,-3$. All yields smaller than $10^{-6} M_\odot$ have been set to zero.}
\label{tab:sndust_initial_rotating}
\begin{tabular}{cllllllll} %modello rotante a metallicità solare massa iniziale tutte densità
$m_\star\,[M_\odot]$	&	$m_\mathrm{dust}\,[M_\odot]$ 	&	$\ce{Al2O3}\,[M_\odot]$	&	$\ce{Fe}\,[M_\odot]$	&	$\ce{Fe3O4}\,[M_\odot]$	&	$\ce{MgSiO3}\,[M_\odot]$	&	$\ce{Mg2SiO4}\,[M_\odot]$	&	$\ce{AC}\,[M_\odot]$	&	$\ce{SiO2}\,[M_\odot]$	\\
\hline
\hline
13	&0.77	&5.1$\times 10^{-3}$	&0	&0.14	&6.7$\times 10^{-5}$	&0.12	&0.028	&0.48	\\
15	&2.2	&7.8$\times 10^{-3}$	&0	&1.2	&3.6$\times 10^{-5}$	&0.17	&2.3$\times 10^{-4}$	&0.84	\\
20	&0.13	&1.9$\times 10^{-3}$	&0	&0.024	&0	&0.056	&0.052	&0	\\
25	&0.15	&0.015	&0	&0.032	&0	&0.092	&8.4$\times 10^{-3}$	&0	\\
30	&0.25	&0.034	&0	&0.038	&0	&0.17	&0	&0	\\
40	&0.34	&0.011	&0	&0.050	&0	&0.12	&0.16	&0	\\
60	&0.20	&0	&3.3$\times 10^{-5}$	&0	&0	&0	&0.20	&0	\\
80	&0.20	&0	&3.7$\times 10^{-4}$	&0	&0	&0	&0.20	&0	\\
120	&0.27	&0	&0.13	&0	&0	&0	&0.13	&0	\\
\hline
\hline
13	&0.12	&0	&0	&0	&1.4$\times 10^{-3}$	&2.8$\times 10^{-4}$	&0.12	&2.1$\times 10^{-5}$	\\
15	&0.60	&3.9$\times 10^{-6}$	&0	&2.4$\times 10^{-4}$	&0.14	&0.070	&0.20	&0.20	\\
20	&0.45	&0.011	&0	&0	&0.016	&6.5$\times 10^{-3}$	&0.41	&0	\\
25	&0.63	&4.9$\times 10^{-3}$	&0	&0	&0.019	&5.2$\times 10^{-3}$	&0.60	&1.3$\times 10^{-6}$	\\
30	&0.068	&5.4$\times 10^{-4}$	&0	&2.7$\times 10^{-3}$	&0	&0.017	&0.048	&0	\\
40	&0.12	&1.0$\times 10^{-3}$	&0	&3.6$\times 10^{-3}$	&0	&0.022	&0.096	&0	\\
60	&0.32	&0	&1.8$\times 10^{-6}$	&0	&0	&0	&0.32	&0	\\
80	&0.53	&0	&1.1$\times 10^{-6}$	&0	&0	&0	&0.53	&0	\\
120	&0.60	&0	&9.6$\times 10^{-6}$	&0	&0	&0	&0.60	&0	\\
\hline
\hline
13	&0.55	&4.5$\times 10^{-6}$	&0	&0.069	&0.10	&0.087	&0.14	&0.16	\\
15	&0.77	&5.7$\times 10^{-5}$	&0	&0.043	&0.14	&0.12	&0.17	&0.30	\\
20	&0.16	&1.9$\times 10^{-5}$	&0	&0	&1.4$\times 10^{-4}$	&3.3$\times 10^{-5}$	&0.16	&1.1$\times 10^{-6}$	\\
25	&2.6$\times 10^{-3}$	&1.7$\times 10^{-4}$	&0	&1.3$\times 10^{-4}$	&0	&1.8$\times 10^{-3}$	&5.3$\times 10^{-4}$	&0	\\
30	&0.017	&1.6$\times 10^{-4}$	&0	&2.7$\times 10^{-4}$	&0	&2.1$\times 10^{-3}$	&0.015	&0	\\
40	&0.15	&9.8$\times 10^{-4}$	&0	&2.0$\times 10^{-5}$	&0	&3.2$\times 10^{-3}$	&0.15	&0	\\
60	&0.34	&1.1$\times 10^{-4}$	&0	&2.0$\times 10^{-5}$	&4.2$\times 10^{-4}$	&2.0$\times 10^{-3}$	&0.34	&0	\\
\hline
\hline
13	&1.3	&4.3$\times 10^{-4}$	&0	&0.67	&0.12	&0.010	&0.34	&0.18	\\
15	&1.3	&5.8$\times 10^{-3}$	&0	&0.41	&0.12	&0.022	&0.30	&0.43	\\
20	&0.029	&3.5$\times 10^{-3}$	&0	&0	&9.9$\times 10^{-4}$	&3.7$\times 10^{-3}$	&0.021	&0	\\
25	&0.14	&0	&0	&0	&2.5$\times 10^{-6}$	&0	&0.14	&0	\\
30	&0.26	&4.8$\times 10^{-4}$	&0	&0	&1.4$\times 10^{-4}$	&4.8$\times 10^{-5}$	&0.26	&1.9$\times 10^{-6}$	\\
40	&0.028	&1.1$\times 10^{-3}$	&0	&0	&0	&1.6$\times 10^{-3}$	&0.026	&0	\\
60	&0.062	&6.8$\times 10^{-4}$	&0	&0	&0	&3.2$\times 10^{-3}$	&0.058	&0	\\
\hline
\hline
\end{tabular}
\end{table*}

%%%%%%%%%%%%%
\begin{landscape}
\begin{table*}[!htbp]
\caption{Effective dust yields for non-rotating models with $\text{[Fe/H]}=0$ and three possible values of the inter-stellar medium density: $n_{\rm ISM} = 0.05, 0.5, 5$ cm$^{-3}$. All yields smaller than $10^{-6} M_\odot$ have been set to zero. $\eta$ is a surviving dust mass fraction.} \label{tab:sndust_final_nonrotating_Z1}
\centering
\begin{tabular}{cllllllllll}
\hline\hline
$m_\star\,[M_\odot]$	&	$m_\mathrm{dust}\,[M_\odot]$ 	&	$\ce{Al2O3}\,[M_\odot]$	&	$\ce{Fe}\,[M_\odot]$	&	$\ce{Fe3O4}\,[M_\odot]$	&	$\ce{MgSiO3}\,[M_\odot]$	&	$\ce{Mg2SiO4}\,[M_\odot]$	&	$\ce{AC}\,[M_\odot]$	&	$\ce{SiO2}\,[M_\odot]$  &   $\eta\,[\%]$  &   $n_{\rm ISM}\,[\mathrm{cm^{-3}}]$\\
\hline
13	&$5.6\times 10^{-6}$	&$0$	&$0$	&$0$	&$0$	&$0$	&$5.6\times 10^{-6}$	&$0$	&$2.1\times 10^{-3}$	&0.05	\\
13	&$0$	&$0$	&$0$	&$0$	&$0$	&$0$	&$0$	&$0$	&$0$	&0.5	\\
13	&$0$	&$0$	&$0$	&$0$	&$0$	&$0$	&$0$	&$0$	&$0$	&5	\\
\hline
15	&$1.0\times 10^{-6}$	&$0$	&$0$	&$0$	&$0$	&$0$	&$1.0\times 10^{-6}$	&$0$	&$2.4\times 10^{-4}$	&0.05	\\
15	&$0$	&$0$	&$0$	&$0$	&$0$	&$0$	&$0$	&$0$	&$0$	&0.5	\\
15	&$0$	&$0$	&$0$	&$0$	&$0$	&$0$	&$0$	&$0$	&$0$	&5	\\
\hline
20	&$0.011$	&$2.3\times 10^{-6}$	&$0$	&$0$	&$0$	&$1.1\times 10^{-6}$	&$0.011$	&$0$	&$1.1$	&0.05	\\
20	&$6.4\times 10^{-3}$	&$0$	&$0$	&$6.2\times 10^{-6}$	&$0$	&$0$	&$6.4\times 10^{-3}$	&$0$	&$0.63$	&0.5	\\
20	&$8.8\times 10^{-4}$	&$0$	&$0$	&$0$	&$0$	&$0$	&$8.8\times 10^{-4}$	&$0$	&$0.087$	&5	\\
\hline
25	&$8.3\times 10^{-3}$	&$1.7\times 10^{-6}$	&$0$	&$2.6\times 10^{-4}$	&$0$	&$6.5\times 10^{-3}$	&$5.5\times 10^{-5}$	&$1.5\times 10^{-3}$	&$0.70$	&0.05	\\
25	&$4.5\times 10^{-4}$	&$1.6\times 10^{-5}$	&$0$	&$0$	&$0$	&$4.0\times 10^{-4}$	&$3.2\times 10^{-5}$	&$3.6\times 10^{-6}$	&$0.038$	&0.5	\\
25	&$1.5\times 10^{-4}$	&$1.6\times 10^{-5}$	&$0$	&$7.9\times 10^{-5}$	&$0$	&$0$	&$1.2\times 10^{-5}$	&$4.4\times 10^{-5}$	&$0.013$	&5	\\
\hline
30	&$1.3\times 10^{-4}$	&$1.3\times 10^{-6}$	&$0$	&$0$	&$0$	&$1.3\times 10^{-4}$	&$1.7\times 10^{-6}$	&$0$	&$0.081$	&0.05	\\
30	&$1.1\times 10^{-5}$	&$5.4\times 10^{-6}$	&$0$	&$2.7\times 10^{-6}$	&$0$	&$3.0\times 10^{-6}$	&$0$	&$0$	&$6.8\times 10^{-3}$	&0.5	\\
30	&$5.7\times 10^{-5}$	&$0$	&$0$	&$1.1\times 10^{-6}$	&$0$	&$5.6\times 10^{-5}$	&$0$	&$0$	&$0.035$	&5	\\
\hline
40	&$1.1\times 10^{-3}$	&$1.6\times 10^{-5}$	&$0$	&$8.0\times 10^{-6}$	&$0$	&$9.7\times 10^{-6}$	&$1.1\times 10^{-3}$	&$0$	&$0.46$	&0.05	\\
40	&$4.5\times 10^{-4}$	&$1.5\times 10^{-5}$	&$0$	&$1.5\times 10^{-5}$	&$0$	&$2.9\times 10^{-4}$	&$1.3\times 10^{-4}$	&$0$	&$0.18$	&0.5	\\
40	&$5.3\times 10^{-4}$	&$0$	&$0$	&$0$	&$0$	&$2.9\times 10^{-4}$	&$2.4\times 10^{-4}$	&$0$	&$0.22$	&5	\\
\hline
60	&$5.0\times 10^{-3}$	&$0$	&$1.4\times 10^{-4}$	&$0$	&$0$	&$0$	&$4.9\times 10^{-3}$	&$0$	&$1.8$	&0.05	\\
60	&$6.5\times 10^{-3}$	&$0$	&$2.5\times 10^{-5}$	&$0$	&$0$	&$0$	&$6.5\times 10^{-3}$	&$0$	&$2.3$	&0.5	\\
60	&$2.0\times 10^{-3}$	&$0$	&$2.6\times 10^{-4}$	&$0$	&$0$	&$0$	&$1.7\times 10^{-3}$	&$0$	&$0.70$	&5	\\
\hline
80	&$7.3\times 10^{-3}$	&$0$	&$2.4\times 10^{-4}$	&$0$	&$0$	&$0$	&$7.1\times 10^{-3}$	&$0$	&$2.2$	&0.05	\\
80	&$7.6\times 10^{-3}$	&$0$	&$3.4\times 10^{-5}$	&$0$	&$0$	&$0$	&$7.6\times 10^{-3}$	&$0$	&$2.3$	&0.5	\\
80	&$3.1\times 10^{-3}$	&$0$	&$2.7\times 10^{-4}$	&$0$	&$0$	&$0$	&$2.8\times 10^{-3}$	&$0$	&$0.91$	&5	\\
\hline
120	&$0.012$	&$0$	&$1.3\times 10^{-3}$	&$0$	&$0$	&$0$	&$0.011$	&$0$	&$2.7$	&0.05	\\
120	&$0.019$	&$0$	&$6.5\times 10^{-5}$	&$0$	&$0$	&$0$	&$0.019$	&$0$	&$4.1$	&0.5	\\
120	&$8.1\times 10^{-3}$	&$0$	&$6.5\times 10^{-4}$	&$0$	&$0$	&$0$	&$7.5\times 10^{-3}$	&$0$	&$1.8$	&5	\\
\hline
\end{tabular}
\end{table*}

\begin{table*}[!htbp]
\caption{Effective dust yields for non-rotating models with $\text{[Fe/H]}=-1$ and three possible values of the interstellar medium density: $n_\mathrm{ISM}=0.05,\,0.5,$ and $5\,\mathrm{cm}^{-3}$. All yields smaller than $10^{-6} M_\odot$ have been set to zero. $\eta$ is a surviving dust mass fraction. }
\label{tab:sndust_final_nonrotating_Z2}
\centering
\begin{tabular}{cllllllllll}
\hline\hline
$m_\star\,[M_\odot]$	&	$m_\mathrm{dust}\,[M_\odot]$ 	&	$\ce{Al2O3}\,[M_\odot]$	&	$\ce{Fe}\,[M_\odot]$	&	$\ce{Fe3O4}\,[M_\odot]$	&	$\ce{MgSiO3}\,[M_\odot]$	&	$\ce{Mg2SiO4}\,[M_\odot]$	&	$\ce{AC}\,[M_\odot]$	&	$\ce{SiO2}\,[M_\odot]$  &   $\eta\,[\%]$  &   $n_{\rm ISM}\,[\mathrm{cm^{-3}}]$\\
\hline
13	&$2.1\times 10^{-6}$	&$0$	&$0$	&$0$	&$0$	&$0$	&$2.1\times 10^{-6}$	&$0$	&$1.9\times 10^{-3}$	&0.05	\\
13	&$0$	&$0$	&$0$	&$0$	&$0$	&$0$	&$0$	&$0$	&$0$	&0.5	\\
13	&$0$	&$0$	&$0$	&$0$	&$0$	&$0$	&$0$	&$0$	&$0$	&5	\\
\hline
15	&$0$	&$0$	&$0$	&$0$	&$0$	&$0$	&$0$	&$0$	&$0$	&0.05	\\
15	&$0$	&$0$	&$0$	&$0$	&$0$	&$0$	&$0$	&$0$	&$0$	&0.5	\\
15	&$0$	&$0$	&$0$	&$0$	&$0$	&$0$	&$0$	&$0$	&$0$	&5	\\
\hline
20	&$9.9\times 10^{-5}$	&$0$	&$0$	&$0$	&$0$	&$0$	&$9.9\times 10^{-5}$	&$0$	&$0.031$	&0.05	\\
20	&$3.9\times 10^{-6}$	&$0$	&$0$	&$0$	&$0$	&$0$	&$3.9\times 10^{-6}$	&$0$	&$1.2\times 10^{-3}$	&0.5	\\
20	&$4.2\times 10^{-6}$	&$0$	&$0$	&$0$	&$0$	&$0$	&$4.2\times 10^{-6}$	&$0$	&$1.3\times 10^{-3}$	&5	\\
\hline
25	&$2.8\times 10^{-3}$	&$0$	&$0$	&$0$	&$0$	&$0$	&$2.8\times 10^{-3}$	&$0$	&$0.83$	&0.05	\\
25	&$3.9\times 10^{-5}$	&$0$	&$0$	&$0$	&$0$	&$0$	&$3.9\times 10^{-5}$	&$0$	&$0.012$	&0.5	\\
25	&$2.9\times 10^{-4}$	&$0$	&$0$	&$0$	&$0$	&$0$	&$2.9\times 10^{-4}$	&$0$	&$0.086$	&5	\\
\hline
30	&$3.6\times 10^{-3}$	&$0$	&$0$	&$0$	&$0$	&$0$	&$3.6\times 10^{-3}$	&$0$	&$1.7$	&0.05	\\
30	&$4.0\times 10^{-3}$	&$0$	&$0$	&$0$	&$0$	&$0$	&$4.0\times 10^{-3}$	&$0$	&$1.9$	&0.5	\\
30	&$7.7\times 10^{-4}$	&$0$	&$0$	&$0$	&$0$	&$0$	&$7.7\times 10^{-4}$	&$0$	&$0.37$	&5	\\
\hline
40	&$7.7\times 10^{-3}$	&$0$	&$0$	&$0$	&$0$	&$0$	&$7.7\times 10^{-3}$	&$0$	&$2.7$	&0.05	\\
40	&$5.2\times 10^{-3}$	&$0$	&$0$	&$0$	&$0$	&$0$	&$5.2\times 10^{-3}$	&$0$	&$1.8$	&0.5	\\
40	&$1.0\times 10^{-3}$	&$0$	&$0$	&$0$	&$0$	&$0$	&$1.0\times 10^{-3}$	&$0$	&$0.34$	&5	\\
\hline
60	&$5.5\times 10^{-3}$	&$0$	&$0$	&$0$	&$0$	&$0$	&$5.5\times 10^{-3}$	&$0$	&$2.5$	&0.05	\\
60	&$3.5\times 10^{-3}$	&$0$	&$0$	&$0$	&$0$	&$0$	&$3.5\times 10^{-3}$	&$0$	&$1.6$	&0.5	\\
60	&$1.3\times 10^{-3}$	&$0$	&$0$	&$0$	&$0$	&$0$	&$1.3\times 10^{-3}$	&$0$	&$0.59$	&5	\\
\hline
80	&$4.7\times 10^{-3}$	&$0$	&$0$	&$1.9\times 10^{-6}$	&$0$	&$7.2\times 10^{-6}$	&$4.7\times 10^{-3}$	&$0$	&$2.7$	&0.05	\\
80	&$9.8\times 10^{-4}$	&$0$	&$0$	&$2.6\times 10^{-6}$	&$0$	&$2.8\times 10^{-6}$	&$9.7\times 10^{-4}$	&$0$	&$0.56$	&0.5	\\
80	&$1.5\times 10^{-3}$	&$0$	&$0$	&$0$	&$0$	&$0$	&$1.5\times 10^{-3}$	&$0$	&$0.85$	&5	\\
\hline
\end{tabular}
\end{table*}

\begin{table*}[!htbp]
\caption{Effective dust yields for non-rotating models with $\text{[Fe/H]}=-2$ and three possible values of the interstellar medium density: $n_\mathrm{ISM}=0.05,\,0.5,$ and $5\,\mathrm{cm}^{-3}$. All yields smaller than $10^{-6} M_\odot$ have been set to zero. $\eta$ is a surviving dust mass fraction. }
\label{tab:sndust_final_nonrotating_Z3}
\centering
\begin{tabular}{cllllllllll}
\hline\hline
$m_\star\,[M_\odot]$	&	$m_\mathrm{dust}\,[M_\odot]$ 	&	$\ce{Al2O3}\,[M_\odot]$	&	$\ce{Fe}\,[M_\odot]$	&	$\ce{Fe3O4}\,[M_\odot]$	&	$\ce{MgSiO3}\,[M_\odot]$	&	$\ce{Mg2SiO4}\,[M_\odot]$	&	$\ce{AC}\,[M_\odot]$	&	$\ce{SiO2}\,[M_\odot]$  &   $\eta\,[\%]$  &   $n_{\rm ISM}\,[\mathrm{cm^{-3}}]$\\
\hline
13	&$0$	&$0$	&$0$	&$0$	&$0$	&$0$	&$0$	&$0$	&$0$	&0.05	\\
13	&$0$	&$0$	&$0$	&$0$	&$0$	&$0$	&$0$	&$0$	&$0$	&0.5	\\
13	&$0$	&$0$	&$0$	&$0$	&$0$	&$0$	&$0$	&$0$	&$0$	&5	\\
\hline
15	&$2.7\times 10^{-5}$	&$0$	&$0$	&$0$	&$0$	&$0$	&$2.7\times 10^{-5}$	&$0$	&$3.2\times 10^{-3}$	&0.05	\\
15	&$0$	&$0$	&$0$	&$0$	&$0$	&$0$	&$0$	&$0$	&$0$	&0.5	\\
15	&$0$	&$0$	&$0$	&$0$	&$0$	&$0$	&$0$	&$0$	&$0$	&5	\\
\hline
20	&$1.1\times 10^{-3}$	&$0$	&$0$	&$0$	&$0$	&$2.0\times 10^{-5}$	&$1.1\times 10^{-3}$	&$0$	&$0.11$	&0.05	\\
20	&$4.4\times 10^{-5}$	&$0$	&$0$	&$0$	&$0$	&$0$	&$4.4\times 10^{-5}$	&$0$	&$4.3\times 10^{-3}$	&0.5	\\
20	&$2.3\times 10^{-4}$	&$0$	&$0$	&$0$	&$0$	&$0$	&$2.3\times 10^{-4}$	&$0$	&$0.022$	&5	\\
\hline
25	&$7.0\times 10^{-3}$	&$0$	&$0$	&$1.3\times 10^{-5}$	&$0$	&$1.7\times 10^{-5}$	&$6.9\times 10^{-3}$	&$3.5\times 10^{-5}$	&$0.60$	&0.05	\\
25	&$6.7\times 10^{-4}$	&$0$	&$0$	&$0$	&$1.2\times 10^{-6}$	&$1.1\times 10^{-5}$	&$6.6\times 10^{-4}$	&$2.3\times 10^{-6}$	&$0.058$	&0.5	\\
25	&$5.9\times 10^{-4}$	&$0$	&$0$	&$0$	&$0$	&$1.3\times 10^{-6}$	&$5.9\times 10^{-4}$	&$0$	&$0.051$	&5	\\
\hline
30	&$4.8\times 10^{-6}$	&$0$	&$0$	&$0$	&$0$	&$1.8\times 10^{-6}$	&$3.0\times 10^{-6}$	&$0$	&$0.035$	&0.05	\\
30	&$3.5\times 10^{-6}$	&$0$	&$0$	&$0$	&$0$	&$0$	&$3.5\times 10^{-6}$	&$0$	&$0.025$	&0.5	\\
30	&$1.5\times 10^{-6}$	&$0$	&$0$	&$0$	&$0$	&$0$	&$1.5\times 10^{-6}$	&$0$	&$0.011$	&5	\\
\hline
40	&$2.5\times 10^{-4}$	&$0$	&$0$	&$0$	&$0$	&$0$	&$2.5\times 10^{-4}$	&$0$	&$1.2$	&0.05	\\
40	&$5.4\times 10^{-4}$	&$0$	&$0$	&$0$	&$0$	&$0$	&$5.4\times 10^{-4}$	&$0$	&$2.6$	&0.5	\\
40	&$1.1\times 10^{-4}$	&$0$	&$0$	&$0$	&$0$	&$0$	&$1.1\times 10^{-4}$	&$0$	&$0.52$	&5	\\
\hline
\end{tabular}
\end{table*}

\begin{table*}[!htbp]
\caption{Effective dust yields for non-rotating models with $\text{[Fe/H]}=-3$ and three possible values of the interstellar medium density: $n_\mathrm{ISM}=0.05,\,0.5,$ and $5\,\mathrm{cm}^{-3}$. All yields smaller than $10^{-6} M_\odot$ have been set to zero. $\eta$ is a surviving dust mass fraction.}
\label{tab:sndust_final_nonrotating_Z4}
\centering
\begin{tabular}{cllllllllll}
\hline\hline
$m_\star\,[M_\odot]$	&	$m_\mathrm{dust}\,[M_\odot]$ 	&	$\ce{Al2O3}\,[M_\odot]$	&	$\ce{Fe}\,[M_\odot]$	&	$\ce{Fe3O4}\,[M_\odot]$	&	$\ce{MgSiO3}\,[M_\odot]$	&	$\ce{Mg2SiO4}\,[M_\odot]$	&	$\ce{AC}\,[M_\odot]$	&	$\ce{SiO2}\,[M_\odot]$  &   $\eta\,[\%]$  &   $n_{\rm ISM}\,[\mathrm{cm^{-3}}]$\\
\hline
13	&$0$	&$0$	&$0$	&$0$	&$0$	&$0$	&$0$	&$0$	&$0$	&0.05	\\
13	&$0$	&$0$	&$0$	&$0$	&$0$	&$0$	&$0$	&$0$	&$0$	&0.5	\\
13	&$0$	&$0$	&$0$	&$0$	&$0$	&$0$	&$0$	&$0$	&$0$	&5	\\
\hline
15	&$1.0\times 10^{-4}$	&$0$	&$0$	&$0$	&$2.0\times 10^{-5}$	&$0$	&$8.1\times 10^{-5}$	&$0$	&$0.016$	&0.05	\\
15	&$0$	&$0$	&$0$	&$0$	&$0$	&$0$	&$0$	&$0$	&$0$	&0.5	\\
15	&$0$	&$0$	&$0$	&$0$	&$0$	&$0$	&$0$	&$0$	&$0$	&5	\\
\hline
20	&$2.4\times 10^{-3}$	&$0$	&$0$	&$0$	&$0$	&$1.3\times 10^{-5}$	&$2.4\times 10^{-3}$	&$1.2\times 10^{-5}$	&$0.22$	&0.05	\\
20	&$2.5\times 10^{-5}$	&$0$	&$0$	&$0$	&$0$	&$0$	&$2.3\times 10^{-5}$	&$2.2\times 10^{-6}$	&$2.3\times 10^{-3}$	&0.5	\\
20	&$2.9\times 10^{-4}$	&$0$	&$0$	&$0$	&$0$	&$0$	&$2.9\times 10^{-4}$	&$0$	&$0.027$	&5	\\
\hline
25	&$4.4\times 10^{-3}$	&$0$	&$0$	&$1.7\times 10^{-5}$	&$1.5\times 10^{-6}$	&$4.0\times 10^{-5}$	&$4.3\times 10^{-3}$	&$4.3\times 10^{-5}$	&$0.36$	&0.05	\\
25	&$2.2\times 10^{-4}$	&$0$	&$0$	&$0$	&$1.6\times 10^{-6}$	&$5.5\times 10^{-6}$	&$2.1\times 10^{-4}$	&$3.3\times 10^{-6}$	&$0.018$	&0.5	\\
25	&$6.9\times 10^{-4}$	&$0$	&$0$	&$0$	&$0$	&$0$	&$6.9\times 10^{-4}$	&$0$	&$0.056$	&5	\\
\hline
30	&$4.9\times 10^{-3}$	&$0$	&$0$	&$0$	&$0$	&$0$	&$4.9\times 10^{-3}$	&$0$	&$2.6$	&0.05	\\
30	&$1.1\times 10^{-3}$	&$0$	&$0$	&$0$	&$0$	&$0$	&$1.1\times 10^{-3}$	&$0$	&$0.58$	&0.5	\\
30	&$2.9\times 10^{-4}$	&$0$	&$0$	&$0$	&$0$	&$0$	&$2.9\times 10^{-4}$	&$0$	&$0.15$	&5	\\
\hline
\end{tabular}
\end{table*}

%%%%%%%%%%%%%%%%%%%%%%%%%%%%%%%%%%%%%%%%%%%%%%%%%%

\begin{table*}[!htbp]
\caption{Effective dust yields for rotating models with $\text{[Fe/H]}=0$ and three possible values of the interstellar medium density: $n_\mathrm{ISM}=0.05,\,0.5,$ and $5\,\mathrm{cm}^{-3}$. All yields smaller than $10^{-6} M_\odot$ have been set to zero. $\eta$ is a surviving dust mass fraction. }
\label{tab:sndust_final_rotating_Z1}
\centering
\begin{tabular}{cllllllllll}
\hline\hline
$m_\star\,[M_\odot]$	&	$m_\mathrm{dust}\,[M_\odot]$ 	&	$\ce{Al2O3}\,[M_\odot]$	&	$\ce{Fe}\,[M_\odot]$	&	$\ce{Fe3O4}\,[M_\odot]$	&	$\ce{MgSiO3}\,[M_\odot]$	&	$\ce{Mg2SiO4}\,[M_\odot]$	&	$\ce{AC}\,[M_\odot]$	&	$\ce{SiO2}\,[M_\odot]$  &   $\eta\,[\%]$  &   $n_{\rm ISM}\,[\mathrm{cm^{-3}}]$\\
\hline
13	&$3.9\times 10^{-3}$	&$0$	&$0$	&$4.0\times 10^{-5}$	&$0$	&$5.9\times 10^{-4}$	&$1.2\times 10^{-4}$	&$3.1\times 10^{-3}$	&$0.50$	&0.05	\\
13	&$3.9\times 10^{-3}$	&$0$	&$0$	&$0$	&$0$	&$0$	&$0$	&$3.9\times 10^{-3}$	&$0.50$	&0.5	\\
13	&$0$	&$0$	&$0$	&$0$	&$0$	&$0$	&$0$	&$0$	&$0$	&5	\\
\hline
15	&$6.1\times 10^{-3}$	&$0$	&$0$	&$2.2\times 10^{-3}$	&$0$	&$1.6\times 10^{-3}$	&$6.0\times 10^{-6}$	&$2.3\times 10^{-3}$	&$0.28$	&0.05	\\
15	&$0.017$	&$0$	&$0$	&$0.010$	&$0$	&$6.9\times 10^{-5}$	&$1.2\times 10^{-6}$	&$7.4\times 10^{-3}$	&$0.79$	&0.5	\\
15	&$1.1\times 10^{-3}$	&$0$	&$0$	&$8.0\times 10^{-6}$	&$0$	&$0$	&$0$	&$1.1\times 10^{-3}$	&$0.050$	&5	\\
\hline
20	&$1.3\times 10^{-3}$	&$0$	&$0$	&$0$	&$0$	&$0$	&$1.3\times 10^{-3}$	&$0$	&$0.97$	&0.05	\\
20	&$5.4\times 10^{-4}$	&$0$	&$0$	&$0$	&$0$	&$0$	&$5.4\times 10^{-4}$	&$0$	&$0.40$	&0.5	\\
20	&$2.9\times 10^{-5}$	&$0$	&$0$	&$0$	&$0$	&$0$	&$2.9\times 10^{-5}$	&$0$	&$0.022$	&5	\\
\hline
25	&$2.5\times 10^{-4}$	&$0$	&$0$	&$0$	&$0$	&$1.4\times 10^{-4}$	&$1.1\times 10^{-4}$	&$0$	&$0.17$	&0.05	\\
25	&$5.6\times 10^{-6}$	&$2.2\times 10^{-6}$	&$0$	&$0$	&$0$	&$1.7\times 10^{-6}$	&$1.7\times 10^{-6}$	&$0$	&$3.8\times 10^{-3}$	&0.5	\\
25	&$1.5\times 10^{-5}$	&$0$	&$0$	&$0$	&$0$	&$4.0\times 10^{-6}$	&$1.1\times 10^{-5}$	&$0$	&$0.010$	&5	\\
\hline
30	&$6.1\times 10^{-4}$	&$2.1\times 10^{-5}$	&$0$	&$1.5\times 10^{-6}$	&$0$	&$5.9\times 10^{-4}$	&$0$	&$0$	&$0.25$	&0.05	\\
30	&$2.6\times 10^{-5}$	&$2.0\times 10^{-5}$	&$0$	&$0$	&$0$	&$5.5\times 10^{-6}$	&$0$	&$0$	&$0.011$	&0.5	\\
30	&$1.3\times 10^{-4}$	&$3.6\times 10^{-5}$	&$0$	&$0$	&$0$	&$9.3\times 10^{-5}$	&$0$	&$0$	&$0.053$	&5	\\
\hline
40	&$3.4\times 10^{-3}$	&$9.6\times 10^{-6}$	&$0$	&$2.6\times 10^{-6}$	&$0$	&$4.0\times 10^{-5}$	&$3.3\times 10^{-3}$	&$0$	&$0.98$	&0.05	\\
40	&$5.7\times 10^{-3}$	&$5.2\times 10^{-6}$	&$0$	&$2.4\times 10^{-5}$	&$0$	&$2.0\times 10^{-4}$	&$5.5\times 10^{-3}$	&$0$	&$1.7$	&0.5	\\
40	&$1.2\times 10^{-3}$	&$0$	&$0$	&$0$	&$0$	&$2.7\times 10^{-4}$	&$9.4\times 10^{-4}$	&$0$	&$0.35$	&5	\\
\hline
60	&$4.0\times 10^{-3}$	&$0$	&$0$	&$0$	&$0$	&$0$	&$4.0\times 10^{-3}$	&$0$	&$2.0$	&0.05	\\
60	&$8.3\times 10^{-3}$	&$0$	&$0$	&$0$	&$0$	&$0$	&$8.3\times 10^{-3}$	&$0$	&$4.1$	&0.5	\\
60	&$1.9\times 10^{-3}$	&$0$	&$0$	&$0$	&$0$	&$0$	&$1.9\times 10^{-3}$	&$0$	&$0.95$	&5	\\
\hline
80	&$6.4\times 10^{-3}$	&$0$	&$1.9\times 10^{-6}$	&$0$	&$0$	&$0$	&$6.4\times 10^{-3}$	&$0$	&$3.2$	&0.05	\\
80	&$9.5\times 10^{-3}$	&$0$	&$0$	&$0$	&$0$	&$0$	&$9.5\times 10^{-3}$	&$0$	&$4.7$	&0.5	\\
80	&$2.4\times 10^{-3}$	&$0$	&$0$	&$0$	&$0$	&$0$	&$2.4\times 10^{-3}$	&$0$	&$1.2$	&5	\\
\hline
120	&$8.2\times 10^{-3}$	&$0$	&$2.6\times 10^{-3}$	&$0$	&$0$	&$0$	&$5.6\times 10^{-3}$	&$0$	&$3.2$	&0.05	\\
120	&$7.6\times 10^{-3}$	&$0$	&$6.1\times 10^{-4}$	&$0$	&$0$	&$0$	&$7.0\times 10^{-3}$	&$0$	&$2.9$	&0.5	\\
120	&$5.2\times 10^{-3}$	&$0$	&$2.5\times 10^{-4}$	&$0$	&$0$	&$0$	&$4.9\times 10^{-3}$	&$0$	&$2.0$	&5	\\
\hline
\end{tabular}
\end{table*}

%%%%%%%%%%%%%%%%%%%%%%%%%%%%%%%%%%%%%%%%%%%%%%%%%%
\begin{table*}[!htbp]
\caption{Effective dust yields for rotating models with $\text{[Fe/H]}=-1$ and three possible values of the interstellar medium density: $n_\mathrm{ISM}=0.05,\,0.5,$ and $5\,\mathrm{cm}^{-3}$. All yields smaller than $10^{-6} M_\odot$ have been set to zero. $\eta$ is a surviving dust mass fraction. }
\label{tab:sndust_final_rotating_Z2}
\centering
\begin{tabular}{cllllllllll}
\hline\hline
$m_\star\,[M_\odot]$	&	$m_\mathrm{dust}\,[M_\odot]$ 	&	$\ce{Al2O3}\,[M_\odot]$	&	$\ce{Fe}\,[M_\odot]$	&	$\ce{Fe3O4}\,[M_\odot]$	&	$\ce{MgSiO3}\,[M_\odot]$	&	$\ce{Mg2SiO4}\,[M_\odot]$	&	$\ce{AC}\,[M_\odot]$	&	$\ce{SiO2}\,[M_\odot]$  &   $\eta\,[\%]$  &   $n_{\rm ISM}\,[\mathrm{cm^{-3}}]$\\
\hline
13	&$8.3\times 10^{-6}$	&$0$	&$0$	&$0$	&$2.1\times 10^{-6}$	&$0$	&$6.2\times 10^{-6}$	&$0$	&$6.8\times 10^{-3}$	&0.05	\\
13	&$2.7\times 10^{-6}$	&$0$	&$0$	&$0$	&$0$	&$0$	&$2.7\times 10^{-6}$	&$0$	&$2.2\times 10^{-3}$	&0.5	\\
13	&$2.2\times 10^{-6}$	&$0$	&$0$	&$0$	&$0$	&$0$	&$2.2\times 10^{-6}$	&$0$	&$1.8\times 10^{-3}$	&5	\\
\hline
15	&$1.9\times 10^{-3}$	&$0$	&$0$	&$0$	&$3.7\times 10^{-4}$	&$7.1\times 10^{-4}$	&$4.7\times 10^{-6}$	&$8.1\times 10^{-4}$	&$0.31$	&0.05	\\
15	&$1.4\times 10^{-3}$	&$0$	&$0$	&$0$	&$2.3\times 10^{-4}$	&$5.3\times 10^{-4}$	&$0$	&$6.1\times 10^{-4}$	&$0.22$	&0.5	\\
15	&$1.6\times 10^{-4}$	&$0$	&$0$	&$0$	&$4.3\times 10^{-5}$	&$5.6\times 10^{-5}$	&$0$	&$6.6\times 10^{-5}$	&$0.027$	&5	\\
\hline
20	&$3.2\times 10^{-3}$	&$0$	&$0$	&$0$	&$4.0\times 10^{-5}$	&$3.5\times 10^{-6}$	&$3.2\times 10^{-3}$	&$0$	&$0.73$	&0.05	\\
20	&$3.2\times 10^{-5}$	&$0$	&$0$	&$0$	&$2.4\times 10^{-5}$	&$2.2\times 10^{-6}$	&$6.0\times 10^{-6}$	&$0$	&$7.3\times 10^{-3}$	&0.5	\\
20	&$9.5\times 10^{-5}$	&$0$	&$0$	&$0$	&$4.2\times 10^{-6}$	&$0$	&$9.1\times 10^{-5}$	&$0$	&$0.021$	&5	\\
\hline
25	&$3.7\times 10^{-3}$	&$0$	&$0$	&$0$	&$1.5\times 10^{-5}$	&$2.3\times 10^{-6}$	&$3.7\times 10^{-3}$	&$0$	&$0.59$	&0.05	\\
25	&$3.9\times 10^{-5}$	&$0$	&$0$	&$0$	&$8.4\times 10^{-6}$	&$1.7\times 10^{-6}$	&$2.9\times 10^{-5}$	&$0$	&$6.2\times 10^{-3}$	&0.5	\\
25	&$5.3\times 10^{-4}$	&$0$	&$0$	&$0$	&$1.1\times 10^{-6}$	&$0$	&$5.3\times 10^{-4}$	&$0$	&$0.084$	&5	\\
\hline
30	&$1.1\times 10^{-3}$	&$0$	&$0$	&$0$	&$0$	&$4.0\times 10^{-6}$	&$1.1\times 10^{-3}$	&$0$	&$1.6$	&0.05	\\
30	&$4.8\times 10^{-4}$	&$0$	&$0$	&$0$	&$0$	&$3.7\times 10^{-6}$	&$4.8\times 10^{-4}$	&$0$	&$0.71$	&0.5	\\
30	&$3.4\times 10^{-5}$	&$0$	&$0$	&$0$	&$0$	&$0$	&$3.4\times 10^{-5}$	&$0$	&$0.050$	&5	\\
\hline
40	&$2.2\times 10^{-3}$	&$0$	&$0$	&$0$	&$0$	&$6.0\times 10^{-6}$	&$2.2\times 10^{-3}$	&$0$	&$1.8$	&0.05	\\
40	&$1.9\times 10^{-3}$	&$0$	&$0$	&$0$	&$0$	&$3.7\times 10^{-6}$	&$1.9\times 10^{-3}$	&$0$	&$1.6$	&0.5	\\
40	&$3.5\times 10^{-4}$	&$0$	&$0$	&$0$	&$0$	&$0$	&$3.5\times 10^{-4}$	&$0$	&$0.29$	&5	\\
\hline
60	&$7.9\times 10^{-3}$	&$0$	&$0$	&$0$	&$0$	&$0$	&$7.9\times 10^{-3}$	&$0$	&$2.5$	&0.05	\\
60	&$0.010$	&$0$	&$0$	&$0$	&$0$	&$0$	&$0.010$	&$0$	&$3.1$	&0.5	\\
60	&$2.0\times 10^{-3}$	&$0$	&$0$	&$0$	&$0$	&$0$	&$2.0\times 10^{-3}$	&$0$	&$0.62$	&5	\\
\hline
80	&$0.013$	&$0$	&$0$	&$0$	&$0$	&$0$	&$0.013$	&$0$	&$2.5$	&0.05	\\
80	&$0.019$	&$0$	&$0$	&$0$	&$0$	&$0$	&$0.019$	&$0$	&$3.6$	&0.5	\\
80	&$4.7\times 10^{-3}$	&$0$	&$0$	&$0$	&$0$	&$0$	&$4.7\times 10^{-3}$	&$0$	&$0.89$	&5	\\
\hline
120	&$0.021$	&$0$	&$0$	&$0$	&$0$	&$0$	&$0.021$	&$0$	&$3.5$	&0.05	\\
120	&$0.029$	&$0$	&$0$	&$0$	&$0$	&$0$	&$0.029$	&$0$	&$4.8$	&0.5	\\
120	&$0.012$	&$0$	&$0$	&$0$	&$0$	&$0$	&$0.012$	&$0$	&$2.0$	&5	\\
\hline
\end{tabular}
\end{table*}

%%%%%%%%%%%%%%%%%%%%%%%%%%%%%%%%%%%%%%%%%%%%%%%%%%
\begin{table*}[!htbp]
\caption{Effective dust yields for rotating models with $\text{[Fe/H]}=-2$ and three possible values of the interstellar medium density: $n_\mathrm{ISM}=0.05,\,0.5,$ and $5\,\mathrm{cm}^{-3}$. All yields smaller than $10^{-6} M_\odot$ have been set to zero. $\eta$ is a surviving dust mass fraction. }
\label{tab:sndust_final_rotating_Z3}
\centering
\begin{tabular}{cllllllllll}
\hline\hline
$m_\star\,[M_\odot]$	&	$m_\mathrm{dust}\,[M_\odot]$ 	&	$\ce{Al2O3}\,[M_\odot]$	&	$\ce{Fe}\,[M_\odot]$	&	$\ce{Fe3O4}\,[M_\odot]$	&	$\ce{MgSiO3}\,[M_\odot]$	&	$\ce{Mg2SiO4}\,[M_\odot]$	&	$\ce{AC}\,[M_\odot]$	&	$\ce{SiO2}\,[M_\odot]$  &   $\eta\,[\%]$  &   $n_{\rm ISM}\,[\mathrm{cm^{-3}}]$\\
\hline
13	&$9.8\times 10^{-4}$	&$0$	&$0$	&$0$	&$1.2\times 10^{-4}$	&$3.7\times 10^{-4}$	&$5.4\times 10^{-6}$	&$4.8\times 10^{-4}$	&$0.18$	&0.05	\\
13	&$6.8\times 10^{-4}$	&$0$	&$0$	&$0$	&$5.6\times 10^{-5}$	&$2.9\times 10^{-4}$	&$0$	&$3.3\times 10^{-4}$	&$0.12$	&0.5	\\
13	&$8.7\times 10^{-5}$	&$0$	&$0$	&$0$	&$1.3\times 10^{-5}$	&$3.8\times 10^{-5}$	&$0$	&$3.6\times 10^{-5}$	&$0.016$	&5	\\
\hline
15	&$3.2\times 10^{-4}$	&$0$	&$0$	&$0$	&$3.6\times 10^{-5}$	&$1.5\times 10^{-4}$	&$7.2\times 10^{-5}$	&$6.0\times 10^{-5}$	&$0.041$	&0.05	\\
15	&$1.7\times 10^{-4}$	&$0$	&$0$	&$0$	&$1.4\times 10^{-5}$	&$1.1\times 10^{-4}$	&$0$	&$4.5\times 10^{-5}$	&$0.022$	&0.5	\\
15	&$3.4\times 10^{-5}$	&$0$	&$0$	&$0$	&$3.4\times 10^{-6}$	&$2.1\times 10^{-5}$	&$0$	&$9.8\times 10^{-6}$	&$4.4\times 10^{-3}$	&5	\\
\hline
20	&$8.2\times 10^{-5}$	&$0$	&$0$	&$0$	&$0$	&$0$	&$8.2\times 10^{-5}$	&$0$	&$0.051$	&0.05	\\
20	&$0$	&$0$	&$0$	&$0$	&$0$	&$0$	&$0$	&$0$	&$0$	&0.5	\\
20	&$0$	&$0$	&$0$	&$0$	&$0$	&$0$	&$0$	&$0$	&$0$	&5	\\
\hline
25	&$9.1\times 10^{-6}$	&$0$	&$0$	&$0$	&$0$	&$9.1\times 10^{-6}$	&$0$	&$0$	&$0.35$	&0.05	\\
25	&$6.3\times 10^{-6}$	&$0$	&$0$	&$0$	&$0$	&$6.3\times 10^{-6}$	&$0$	&$0$	&$0.24$	&0.5	\\
25	&$0$	&$0$	&$0$	&$0$	&$0$	&$0$	&$0$	&$0$	&$0$	&5	\\
\hline
30	&$2.8\times 10^{-4}$	&$0$	&$0$	&$0$	&$0$	&$1.3\times 10^{-5}$	&$2.7\times 10^{-4}$	&$0$	&$1.6$	&0.05	\\
30	&$1.3\times 10^{-4}$	&$0$	&$0$	&$0$	&$0$	&$9.2\times 10^{-6}$	&$1.2\times 10^{-4}$	&$0$	&$0.74$	&0.5	\\
30	&$7.1\times 10^{-6}$	&$0$	&$0$	&$0$	&$0$	&$1.4\times 10^{-6}$	&$5.7\times 10^{-6}$	&$0$	&$0.041$	&5	\\
\hline
40	&$2.3\times 10^{-3}$	&$0$	&$0$	&$0$	&$0$	&$6.5\times 10^{-6}$	&$2.3\times 10^{-3}$	&$0$	&$1.5$	&0.05	\\
40	&$2.7\times 10^{-3}$	&$0$	&$0$	&$0$	&$0$	&$5.4\times 10^{-6}$	&$2.7\times 10^{-3}$	&$0$	&$1.8$	&0.5	\\
40	&$4.6\times 10^{-4}$	&$0$	&$0$	&$0$	&$0$	&$1.1\times 10^{-6}$	&$4.6\times 10^{-4}$	&$0$	&$0.30$	&5	\\
\hline
60	&$4.9\times 10^{-3}$	&$0$	&$0$	&$0$	&$8.4\times 10^{-6}$	&$2.0\times 10^{-6}$	&$4.9\times 10^{-3}$	&$0$	&$1.4$	&0.05	\\
60	&$8.2\times 10^{-3}$	&$0$	&$0$	&$0$	&$4.2\times 10^{-6}$	&$2.4\times 10^{-6}$	&$8.2\times 10^{-3}$	&$0$	&$2.4$	&0.5	\\
60	&$1.6\times 10^{-3}$	&$0$	&$0$	&$0$	&$0$	&$0$	&$1.6\times 10^{-3}$	&$0$	&$0.47$	&5	\\
\hline
\end{tabular}
\end{table*}

%%%%%%%%%%%%%%%%%%%%%%%%%%%%%%%%%%%%%%%%%%%%%%%%%%
\begin{table*}[!htbp]
\caption{Effective dust yields for rotating models with  $\text{[Fe/H]}=-3$ and three possible values of the interstellar medium density: $n_\mathrm{ISM}=0.05,\,0.5,$ and $5\,\mathrm{cm}^{-3}$. All yields smaller than $10^{-6} M_\odot$ have been set to zero. $\eta$ is a surviving dust mass fraction. }
\label{tab:sndust_final_rotating_Z4}
\centering
\begin{tabular}{cllllllllll}
\hline\hline
$m_\star\,[M_\odot]$	&	$m_\mathrm{dust}\,[M_\odot]$ 	&	$\ce{Al2O3}\,[M_\odot]$	&	$\ce{Fe}\,[M_\odot]$	&	$\ce{Fe3O4}\,[M_\odot]$	&	$\ce{MgSiO3}\,[M_\odot]$	&	$\ce{Mg2SiO4}\,[M_\odot]$	&	$\ce{AC}\,[M_\odot]$	&	$\ce{SiO2}\,[M_\odot]$  &   $\eta\,[\%]$  &   $n_{\rm ISM}\,[\mathrm{cm^{-3}}]$\\
\hline
13	&$3.2\times 10^{-3}$	&$0$	&$0$	&$5.1\times 10^{-5}$	&$1.7\times 10^{-3}$	&$9.5\times 10^{-5}$	&$1.5\times 10^{-4}$	&$1.2\times 10^{-3}$	&$0.24$	&0.05	\\
13	&$2.2\times 10^{-3}$	&$0$	&$0$	&$2.3\times 10^{-5}$	&$1.2\times 10^{-3}$	&$8.3\times 10^{-5}$	&$0$	&$8.9\times 10^{-4}$	&$0.17$	&0.5	\\
13	&$3.1\times 10^{-4}$	&$0$	&$0$	&$4.8\times 10^{-6}$	&$2.1\times 10^{-4}$	&$8.9\times 10^{-6}$	&$0$	&$8.2\times 10^{-5}$	&$0.023$	&5	\\
\hline
15	&$3.5\times 10^{-3}$	&$1.1\times 10^{-6}$	&$0$	&$2.2\times 10^{-6}$	&$8.8\times 10^{-5}$	&$8.5\times 10^{-5}$	&$3.3\times 10^{-3}$	&$6.0\times 10^{-6}$	&$0.27$	&0.05	\\
15	&$1.7\times 10^{-4}$	&$0$	&$0$	&$2.0\times 10^{-6}$	&$4.5\times 10^{-5}$	&$7.0\times 10^{-5}$	&$4.2\times 10^{-5}$	&$6.5\times 10^{-6}$	&$0.013$	&0.5	\\
15	&$2.2\times 10^{-5}$	&$0$	&$0$	&$0$	&$9.3\times 10^{-6}$	&$1.0\times 10^{-5}$	&$0$	&$2.3\times 10^{-6}$	&$1.7\times 10^{-3}$	&5	\\
\hline
20	&$4.3\times 10^{-5}$	&$0$	&$0$	&$0$	&$1.7\times 10^{-5}$	&$2.9\times 10^{-6}$	&$2.3\times 10^{-5}$	&$0$	&$0.15$	&0.05	\\
20	&$1.5\times 10^{-5}$	&$0$	&$0$	&$0$	&$1.3\times 10^{-5}$	&$1.9\times 10^{-6}$	&$0$	&$0$	&$0.051$	&0.5	\\
20	&$3.3\times 10^{-6}$	&$0$	&$0$	&$0$	&$2.2\times 10^{-6}$	&$0$	&$1.1\times 10^{-6}$	&$0$	&$0.011$	&5	\\
\hline
25	&$3.4\times 10^{-3}$	&$0$	&$0$	&$0$	&$0$	&$0$	&$3.4\times 10^{-3}$	&$0$	&$2.4$	&0.05	\\
25	&$7.9\times 10^{-4}$	&$0$	&$0$	&$0$	&$0$	&$0$	&$7.9\times 10^{-4}$	&$0$	&$0.56$	&0.5	\\
25	&$2.5\times 10^{-5}$	&$0$	&$0$	&$0$	&$0$	&$0$	&$2.5\times 10^{-5}$	&$0$	&$0.018$	&5	\\
\hline
30	&$6.8\times 10^{-3}$	&$0$	&$0$	&$0$	&$0$	&$0$	&$6.8\times 10^{-3}$	&$0$	&$2.6$	&0.05	\\
30	&$3.6\times 10^{-3}$	&$0$	&$0$	&$0$	&$0$	&$0$	&$3.6\times 10^{-3}$	&$0$	&$1.4$	&0.5	\\
30	&$3.1\times 10^{-4}$	&$0$	&$0$	&$0$	&$0$	&$0$	&$3.1\times 10^{-4}$	&$0$	&$0.12$	&5	\\
\hline
40	&$4.1\times 10^{-4}$	&$0$	&$0$	&$0$	&$0$	&$0$	&$4.1\times 10^{-4}$	&$0$	&$1.4$	&0.05	\\
40	&$7.2\times 10^{-5}$	&$0$	&$0$	&$0$	&$0$	&$0$	&$7.2\times 10^{-5}$	&$0$	&$0.25$	&0.5	\\
40	&$3.7\times 10^{-6}$	&$0$	&$0$	&$0$	&$0$	&$0$	&$3.7\times 10^{-6}$	&$0$	&$0.013$	&5	\\
\hline
60	&$1.3\times 10^{-3}$	&$0$	&$0$	&$0$	&$0$	&$7.5\times 10^{-6}$	&$1.3\times 10^{-3}$	&$0$	&$2.1$	&0.05	\\
60	&$3.1\times 10^{-4}$	&$0$	&$0$	&$0$	&$0$	&$7.7\times 10^{-6}$	&$3.0\times 10^{-4}$	&$0$	&$0.50$	&0.5	\\
60	&$1.1\times 10^{-4}$	&$0$	&$0$	&$0$	&$0$	&$1.8\times 10^{-6}$	&$1.1\times 10^{-4}$	&$0$	&$0.18$	&5	\\
\hline
\end{tabular}
\end{table*}
\end{landscape}

\end{appendix}

\end{document}